\documentclass[eqsecnum,floats,aps,nofootinbib,showpacs]{revtex4}
\usepackage{latexsym}
\usepackage{amssymb}

\usepackage[T1]{fontenc}
\usepackage{bm,exscale,latexsym, amsmath, amssymb, mathrsfs, dsfont,accents,tensor} 
\usepackage[dvips]{graphicx}

\renewcommand{\l}{\left(}
\renewcommand{\r}{\right)}
\usepackage{amsmath, amsthm, amssymb}

\def\be{\begin{equation}}
\def\ee{\end{equation}}
\def\beq{\begin{equation*}}
\def\eeq{\end{equation*}}
\def\ba{\begin{aligned}}
\def\ea{\end{aligned}}

\def\ov{\overline}

\begin{document}
\title{Generalized Dirac bracket and the role of the Poincar\'e symmetry \break in the program of canonical quantization of fields 2}

\author {Marcin Ka\'zmierczak}
\email{marcin.kazmierczak@fuw.edu.pl}
\affiliation{Institute of Theoretical Physics, University of Warsaw
  ul. Ho\.{z}a 69, 00-681 Warszawa, Poland}

\begin{abstract}
In this article the methods of canonical analysis and quantization that were reviewed in the first part of the series are applied to the case of the Dirac field in the presence of electromagnetic interaction. It is shown that the quantization of electrodynamics, which begins with a given Lagrangian and ends by perturbative calculation of scattering probability amplitudes, can be performed in the way that does not employ Poincar\'e symmetry of space--time at any stage. Also, the causal structure is not needed.
\end{abstract}
 

\maketitle

\section{Introduction}
In this second part of the sequel, I shall apply the canonical formalism reviewed in the first part to the case of electrodynamics with spinorial matter. I shall begin with the discussion of constraints and gauge transformations in Section \ref{general}. The issue of integrating infinitesimal gauge transformations to the finite form will be addressed and the total and extended Hamiltonian formalisms confronted.

In Section \ref{Coul}, the issue of consistent imposition of gauge conditions will be discussed on the simplest example of Coulomb gauge. The generalized Dirac bracket (GDB) will be constructed for this case. Then equations of motion will be discussed. Since they will appear to be by far too complicated to be exactly solved, the transition to the interaction picture will be necessary. When this is done, the equations simplify immensely and straightforward Fock quantization can be applied to the interaction picture fields. Knowledge of this interaction picture representation will appear to be sufficient for the perturbative description of scattering processes.

In Section \ref{Exam}, the two examples of Compton scattering and electron, positron $\longrightarrow$ muon, anti--muon scattering will be discussed. The elements of the $S$ matrix in the lowest non--trivial order (i.e. the second order in fine structure constant) will be explicitly computed. No Feynman rules will be postulated, although the relation of the calculations to the standard quantization based on Feynman diagrams will be explained.

\section{The structure of constraints of electrodynamics}\label{general}

The theory of electromagnetic interaction is interpreted as a  gauge theory of a $U(1)$ group. The Lagrangian for fermions in the presence of this interaction is obtained through the minimal coupling procedure. Explicitly,
\be
\begin{aligned}
&\mathcal{L}=\frac{i}{2}\l{\overline{\psi}\gamma^a\mathcal{D}_a\psi-\overline{\mathcal{D}_a\psi}\gamma^a\psi}\r
-m\overline{\psi}\psi-\frac{1}{4}F_{ab}F^{ab} 
=\mathcal{L}_D+\mathcal{L}_{EM}-eA_a\overline{\psi}\gamma^a\psi , \\
&\mathcal{L}_D=\frac{i}{2}\l{\overline{\psi}\gamma^a\partial_a\psi-\partial_a\overline{\psi}\gamma^a\psi}\r,\quad
\mathcal{L}_{EM}=-\frac{1}{4}F_{ab}F^{ab},\quad
F_{ab}=\partial_aA_b-\partial_bA_a,\quad,
\mathcal{D}_a\psi:=(\partial_a+ieA_a)\psi ,
\end{aligned}
\ee
where $e$ is the electric charge and $m$ represents the particle's mass. In order for the covariant derivative $\mathcal{D}_a\psi$ to transform as  
$\mathcal{D}_a\psi\rightarrow\mathcal{D}'_a\psi'=e^{-ie\lambda}\mathcal{D}_a\psi$ under the $U(1)$ gauge transformation $\psi\rightarrow\psi'=e^{-ie\lambda}\psi$, the $U(1)$ connection one--form (electromagnetic four--potential) needs to transform as $A_a\rightarrow A'_a=A_a+\partial_a\lambda$. Clearly, the Lagrangian is then invariant under these transformations. The fields of the theory are $\psi$, $\ov{\psi}$ and $A_a$. The equations for conjugated momenta are
\be
\pi=\frac{\partial\mathcal{L}}{\partial(\partial_0\psi)}=\frac{i}{2}\ov{\psi}\gamma^0,\quad
\ov{\pi}=\frac{\partial\mathcal{L}}{\partial(\partial_0\ov{\psi})}=-\frac{i}{2}\gamma^0\psi,\quad
\pi^a=\frac{\partial\mathcal{L}}{\partial(\partial_0 A_a)}=-F^{0a} ,
\ee
from which the primary constraints follow
\be\label{primaryQED}
\chi_1=\pi-\frac{i}{2}\overline{\psi}\gamma^0,\quad \chi_2=\ov{\pi}+\frac{i}{2}\gamma^0\psi,\quad
\gamma_1=\pi^0 .
\ee
There is a possible source of confusion in the notation, since the latter $\gamma$ is used to denote the Dirac matrices, as well as first class constraints (I will argue that $\gamma_1$ is first class later on). The confusion can be avoided if one remembers that the Dirac matrices are denoted by $\gamma$ with superscripts and the constraints are labeled by subscripts.
The canonical Hamiltonian is given by
\be\label{H}
\ba
&H=\int\mathcal{H} d^3x, \\
&\mathcal{H}=\pi\dot{\psi}+\dot{\overline{\psi}}\ov{\pi}+\pi^a\dot{A_a}-\mathcal{L}\\
&=\ov{\psi}\l{-i\gamma^j\partial_j+m}\r\psi+\frac{1}{2}\pi^j\pi^j+\frac{1}{4}F_{ij}F_{ij}+eA_j\ov{\psi}\gamma^j\psi+
A_0\l{e\ov{\psi}\gamma^0\psi-\partial_j\pi^j}\r+\partial_j\l{A_0\pi^j+\frac{i}{2}\ov{\psi}\gamma^j\psi}\r 
\ea
\ee
and the total Hamiltonian is
\be
H_T=H+\int\l{u^0(\vec{x})\gamma_1(\vec{x})+\chi_1(\vec{x})u^1(\vec{x})+u^2(\vec{x})\chi_2(\vec{x})}\r, d^3x
\ee
where $u's$ are arbitrary multipliers (note that $u^1$ and $u^2$ are matrices).
The last component of $\mathcal{H}$ in (\ref{H}) is a three--divergence that results in a surface term in $H$. Such terms do not contribute to the functional derivatives of $H$ with respect to fields, since it is common to consider variations that vanish on the boundary of the integration region when defining variational derivatives. If the fields themselves (not only their variations) vanish at spatial infinity, which is usually assumed in the case of flat space--times, then the surface terms simply vanish thus not giving any contribution to the total energy.

(III.6)

The GPB is given by
\be\label{GPBQED}
[F,G]_{GP}=\int \left({
\frac{\delta F}{\delta A_a(\vec{x})}\frac{\delta G}{\delta \pi^a(\vec{x})}-
\frac{\delta F}{\delta \pi^a(\vec{x})}\frac{\delta G}{\delta A_a(\vec{x})}+
\frac{\delta F}{\delta\psi(\vec{x})} \frac{\delta G}{\delta\pi(\vec{x})} \pm 
\frac{\delta G}{\delta\overline{\pi}(\vec{x})} \frac{\delta F}{\delta\overline{\psi}(\vec{x})} \mp
\frac{\delta G}{\delta\psi(\vec{x})} \frac{\delta F}{\delta\pi(\vec{x})} -
\frac{\delta F}{\delta\overline{\pi}(\vec{x})} \frac{\delta G}{\delta\overline{\psi}(\vec{x})}
}\right) d^3x ,
\ee
where the upper sign applies whenever at least one of the variables $F$, $G$ is even and the lower one corresponds to $F$ and $G$ odd (see formula (III.6) of \cite{Kazm6} and the discussion therein). The consistency conditions for the time evolution of constraints are 
\be\label{consist}
\begin{aligned}
&[\gamma_1,H_T]_{GP}=-\frac{\delta H_T}{\delta A_0}=-\frac{\delta H_T}{\delta A_0}=
\partial_j\pi^j-e\ov{\psi}\gamma^0\psi\approx 0,\\
&[\chi_1,H_T]_{GP}=-\frac{\delta H_T}{\delta \psi}-\frac{i}{2}\frac{\delta H_T}{\delta \ov{\pi}}\gamma^0=
-i\ov{\mathcal{D}_j\psi}\gamma^j-m\ov{\psi}-eA_0\ov{\psi}\gamma^0-iu^2\gamma^0\approx 0, \\
&[\chi_2,H_T]_{GP}=-\frac{\delta H_T}{\delta \ov{\psi}}+\frac{i}{2}\gamma^0\frac{\delta H_T}{\delta \pi}=
i\gamma^j\mathcal{D}_j\psi-m\psi-eA_0\gamma^0\psi+i\gamma^0u^1\approx 0 .
\end{aligned}
\ee
The first equation does not depend on $u's$ and hence gives rise to a secondary constraint
\be
\tilde{\gamma}_2:=\partial_j\pi^j-e\ov{\psi}\gamma^0\psi .
\ee
The second and the third equation of (\ref{consist}) yield merely the restrictions on $u's$ of the form
\be\label{U1U2}
U^1=\l{-\gamma^0\gamma^j\mathcal{D}_j-im\gamma^0}\r\psi-ieA_0\psi,\qquad
U^2=\ov{U^1}=-\ov{\mathcal{D}_j\psi}\gamma^j\gamma^0+im\ov{\psi}\gamma^0+ieA_0\ov{\psi} .
\ee
The bracket of $\tilde{\gamma}_2$ with $H_T$ is 
\be
[\tilde{\gamma}_2,H_T]_{GP}=-e\left[{\partial_j\l{\ov{\psi}\gamma^j\psi}\r+\ov{\psi}\gamma^0u^1+u^2\gamma^0\psi}\right] .
\ee
This is yet another restriction on $u's$, but it is easy to verify that it is satisfied automatically if $u^1=U^1$ and $u^2=U^2$ (see (\ref{U1U2})). Hence, neither farther restrictions on $u's$ nor additional constraints are produced. The consistency algorithm is accomplished.

Let us now investigate what class do the constraints belong to. Clearly $\gamma^1$ commutes with all the others and hence is first class. The brackets of the remaining constraints are nontrivial
\be
\ba
&[\tilde{\gamma}_2(\vec{x}),\chi_{1l\vec{x}'}]_{GP}=-e\l{\ov{\psi}(\vec{x})\gamma^0}\r_l\delta(\vec{x}-\vec{x}'),\quad
[\tilde{\gamma}_2(\vec{x}),\chi_{2l\vec{x}'}]_{GP}=
-e\l{\gamma^0{\psi}(\vec{x})}\r_l\delta(\vec{x}-\vec{x}'),\\
&[\chi_{1l\vec{x}},\chi_{2l'\vec{x}'}]_{GP}=i\gamma^0_{l'l}\delta(\vec{x}-\vec{x}') .
\ea
\ee
It seems that there are three families of second class constraints, $\chi_1$, $\chi_2$, $\tilde{\gamma}$ and one family of first class ones, $\gamma_1$. However, as explained at the end of Sec.II.A of \cite{Kazm6}, the constraints are well separated if the number of first class constraints is possibly large, i.e. no first class constraints are hidden in the second class ones.  Are these constraints well separated? To verify this, let us try to construct a first class constraint $\gamma$ from a linear combination of $\chi_1$'s, $\chi_2$'s and $\tilde{\gamma}$'s
\be
\gamma:=\int
\l{\chi_1(\vec{x})\lambda_1(\vec{x})+\lambda_2(\vec{x})\chi_2(\vec{x})+
\kappa(\vec{x})\tilde{\gamma}_2(\vec{x})}\r d^3x ,
\ee
where $\lambda$'s are matrices and $\kappa$ a scalar function. In order for this constraint to commute with all the others, it is necessary the the following relations between $\lambda_1$, $\lambda_2$ and $\kappa$ hold
\be
\ov{\psi}\gamma^0\lambda_1\approx-\lambda_2\gamma^0\psi,\quad
\lambda_2\approx-ie\ov{\psi},\quad
\lambda_1\approx ie\kappa\psi .
\ee
They are sufficient for $\gamma$ to be first class. Hence, a first class constraint
\be
\gamma=\int\kappa(\vec{x})
\l{\tilde{\gamma}_2(\vec{x})+ie\chi_1(\vec{x})\psi(\vec{x})-ie\ov{\psi}(\vec{x})\chi_2(\vec{x})}\r d^3x
\ee
can be constructed. Since $\kappa$ remained completely arbitrary, the family of first class constraints 
\be
\gamma_2(\vec{x}):=\tilde{\gamma}_2(\vec{x})+ie\chi_1(\vec{x})\psi(\vec{x})-ie\ov{\psi}(\vec{x})\chi_2(\vec{x})
\ee
was constructed. Since the system of constraints $\chi_1$, $\chi_2$, $\gamma_1$, $\tilde{\gamma}_2$ is equivalent to $\chi_1$, $\chi_2$, $\gamma_1$, $\gamma_2$, we are allowed to use the latter. To summarize, the constraints of the theory are
\be
\gamma_1=\pi^0,\qquad \gamma_2=\partial_j\pi^j+ie(\pi\psi-\ov{\psi}\ov{\pi}),\qquad
\chi_1=\pi-\frac{i}{2}\ov{\psi}\gamma^0,\qquad \chi_2=\ov{\pi}+\frac{i}{2}\gamma^0\psi ,
\ee
where $\gamma$'s are first class and $\chi$'s are second class. These constraints are well separated, since it is impossible to construct a first class constraint from linear combination of $\chi_1$'s and $\chi_2$'s.
Using (\ref{U1U2}), the first class Hamiltonian $H'$ can be calculated
\be
\ba
&H'=H+\int\l{\chi_1 U^1+U^2\chi_2}\r d^3x \\
&=\int\left[{\ov{\psi}\l{-i\gamma^j\mathcal{D}_j+m}\r\psi+\frac{1}{2}\pi^j\pi^j+\frac{1}{4}F_{ij}F_{ij}-
A_0\gamma_2+\chi_1\l{-\gamma^0\gamma^j\mathcal{D}_j\psi-im\gamma^0\psi}\r+
\l{-\ov{\mathcal{D}_j\psi}\gamma^j\gamma^0+im\ov{\psi}\gamma^0}\r\chi_2}\right] d^3x .
\ea
\ee
The total and extended Hamiltonians are then
\be\label{HTHE}
H_T=H'+\int u^0(\vec{x})\gamma_1(\vec{x}) d^3 x,\qquad
H_E=H'+\int \l{u^0(\vec{x})\gamma_1(\vec{x})+w(\vec{x})\gamma_2(\vec{x})}\r d^3x ,
\ee
where $u^0$ and $w$ are arbitrary functions.

\subsection{Equations of motion}

 The equation of motion of any dynamical variable $F$ generated by $H_E$ is $\dot{F}=[F,H_E]_{GP}$. For basic canonical variables this gives
\be\label{emQED}
\begin{aligned}
&\dot{A}_0=u_0, & (\ref{emQED}a) \\
&\dot{A}_j=\pi^j+\partial_j(A_0-w), & (\ref{emQED}b) \\
&\dot{\pi}^0=\partial_j\pi^j+ie(\pi\psi-\ov{\psi}\ov{\pi})=\gamma_2, & (\ref{emQED}c) \\
&\dot{\pi}^j=\partial_iF_{ij}+ie(\pi\gamma^0\gamma^j\psi+\ov{\psi}\gamma^0\gamma^j\ov{\pi}), & (\ref{emQED}d) \\
&\dot{\psi}=\l{-\gamma^0\gamma^j\mathcal{D}_j-im\gamma^0}\r\psi-ie(A_0-w)\psi, & (\ref{emQED}e) \\
&\dot{\ov{\psi}}=-\ov{\mathcal{D}_j\psi}\gamma^j\gamma^0+im\ov{\psi}\gamma^0+ie(A_0-w)\ov{\psi}=\ov{\dot{\psi}} , & (\ref{emQED}f) \\
&\dot{\pi}=-\mathcal{D}_j\pi\gamma^0\gamma^j+im\pi\gamma^0+ie(A_0-w)\pi, & (\ref{emQED}g) \\
&\dot{\ov{\pi}}=
\l{-\gamma^0\gamma^j\mathcal{D}_j-im\gamma^0}\r\ov{\pi}-ie(A_0-w)\ov{\pi}=\ov{\dot{\pi}},  & (\ref{emQED}h) \\
&\pi^0=0, & (\ref{emQED}i) \\
&\partial_j\pi^j=ie(\ov{\psi}\ov{\pi}-\pi\psi), & (\ref{emQED}j) \\
&\pi=\frac{i}{2}\ov{\psi}\gamma^0, & (\ref{emQED}k) \\
&\ov{\pi}=-\frac{i}{2}\gamma^0\psi, & (\ref{emQED}l) \\
&\mathcal{D}_j\psi:=(\partial_j+ieA_j)\psi,\quad \mathcal{D}_j\ov{\pi}:=(\partial_j+ieA_j)\ov{\pi},\quad
\mathcal{D}_j\ov{\psi}:=\ov{\mathcal{D}_j\psi},\quad \mathcal{D}_j\pi:=\ov{\mathcal{D}_j\ov{\pi}}.
\end{aligned}
\ee
The first four equations $a,b,c,d,e,f,g,h$ follow from evaluation of GP of the variables with $H_E$ and the equations $i,j,k,l$ are the constraints. A high degree of redundancy can be seen in this system. The equations $i,k,l$ and $b$ can be used to eliminate $\pi^0$, $\pi$, $\ov{\pi}$ and $\pi^j$ from the remaining ones. Then (\ref{emQED}a) just states that $A_0$ is an arbitrary function,  (\ref{emQED}c) becomes equivalent to  (\ref{emQED}j), 
 (\ref{emQED}d) gives
\be\label{Beq}
\partial_aF^{aj}=e\ov{\psi}\gamma^j\psi+\partial_j\dot{w},
\ee
(\ref{emQED}e) gives
\be\label{Deq}
\l{i\gamma^a\mathcal{D}_a-m}\r\psi=-ew\gamma^0\psi,
\ee
 (\ref{emQED}f) follows from (\ref{emQED}e) by Dirac conjugation, (\ref{emQED}g) and (\ref{emQED}h) reduce to (\ref{emQED}f) and (\ref{emQED}e) on the constraint surface. Finally, after elimination of $\pi$, $\ov{\pi}$ and $\pi^j$, (\ref{emQED}j) gives
\be\label{Eeq}
\partial_aF^{a0}=e\ov{\psi}\gamma^0\psi-\partial_i\partial_iw .
\ee

In the case when $w=0$, (\ref{Beq}) and (\ref{Eeq}) are Maxwell equations and (\ref{Eeq}) reduce to the $U(1)$--covariant Dirac equation of spinor electrodynamics. In the $w\not= 0$ case the equations, when written in terms of $A_a$, do not assume the usual form of Maxwell equations. This example reflects a general rule that it is the dynamics generated by $H_T$, and not $H_E$, which is equivalent to the Euler--Lagrange equations of the initial Lagrangian (see the discussion below the formula (II.20) of \cite{Kazm6}) . However, if $A_0$ is transformed into $\tilde{A_0}:=A_0-w$ then, in terms of the quantities $\tilde{F}_{ab}$ and $\tilde{\mathcal{D}}\psi$ that contain $\tilde{A_0}$ in place of $A_0$, the equations assume the standard form
\be
\partial_aF^{ab}=e\ov{\psi}\gamma^a\psi,\qquad \l{i\gamma^a\mathcal{D}_a-m}\r\psi=0 .
\ee
Whether these equations are equivalent to Maxwell equations or not depends on how the physically measurable electric and magnetic fields are defined in terms of $A_a$. The correct definition in the extended formalism is
\be
E^i:=\pi^i=F^{i0}+\partial_iw=\tilde{F}^{i0},\qquad B^i:=\varepsilon^{ijk}\partial_jA^k,
\ee
where $\varepsilon^{ijk}$ is antisymmetric symbol with $\varepsilon^{123}=1$. Note that the position of spatial indexes is important, since in the metric convention $(+,-,-,-)$, which is used, shifting a spatial index leads to the change of sign.

\subsection{Gauge transformations}
The time evolution of any dynamical variable $F$ is given by
\be\label{FevE}
\dot{F}=[F,H_E]_{GP}=
[F,H']_{GP}+\int \l{u^0(\vec{x})[F,\gamma_1(\vec{x})]_{GP}+w(\vec{x})[F,\gamma_2(\vec{x})]_{GP}}\r d^3x
\ee
and depends on arbitrary functions $u^0$ and $w$. The transformations corresponding to changes in these functions ought to be interpreted as gauge transformations, as explained in Sec.II.B of \cite{Kazm6}. From (\ref{FevE}) it is clear that the action of a general such transformation on  $\dot{F}$ is
\be\label{dotFch}
\delta\dot{F}=\int \l{\delta u^0(\vec{x})[F,\gamma_1(\vec{x})]_{GP}+\delta w(\vec{x})[F,\gamma_2(\vec{x})]_{GP}}\r d^3x .
\ee
The transformations corresponding to the change in $u^0$ are said to be generated by the constraint $\gamma_1$, whereas those following from the change in $w$ are generated by $\gamma_2$. Imagine that $F(t)$ and $\tilde{F}(t)$ correspond to the dynamics obtained for the choice of arbitrary functions as $u^0$, $w$ and $\tilde{u^0}=u^0+\delta u^0$, $\tilde{w}=w+\delta w$ respectively. Assume that $F(t_0)=\tilde{F}(t_0)$ at some instant $t_0$. The difference $\delta{F}(t)=\tilde{F}(t)-F(t)$ corresponds to the unphysical gauge transformation. But how can this difference be calculated explicitly? If $t=t_0+\tau$ and $F$ is analytic, then 
\be
\delta{F}(t_0+\tau)=\sum_{n=1}^{\infty}\delta F^{(n)}(t_0)\frac{\tau^n}{n!} ,
\ee 
where $F^{(n)}$ denotes the $n$--th time derivative of $F$. Up to first order in $\tau$, this change can be easily evaluated by means of (\ref{dotFch}). But what if we wished to find a finite form of a gauge transformation?

The equations of motion in the form (\ref{emQED}) are very helpful. First note that instead of using (\ref{dotFch}) it is straightforward to obtain the variations of first derivatives of basic canonical fields directly from (\ref{emQED}). From now on, we will be interested in changes in the derivatives at $t=t_0$ but the argument $t_0$ will not be written explicitly. So for example $\delta F:=\tilde{F}(t_0)-F(t_0)=0$ for $F$ being any canonical variable, but $\delta\dot{F}:=\dot{\tilde{F}}(t_0)-\dot{F}(t_0)$ will not vanish in general. Specifically, from (\ref{emQED}) it follows that
\be
\delta \dot{A_0}=\delta u^0,\quad \delta\dot{A}_j=-\partial_j\delta w,\quad \delta\dot{\pi}^0=\delta\dot{\pi}^j=0,\quad \delta\dot{\psi}=ie\delta w\psi,\quad
\delta\dot{\ov{\psi}}=-ie\delta w\ov{\psi},\quad
\delta\dot{\pi}=-ie\delta w\pi,\quad \delta\dot{\ov{\pi}}=ie\delta w\ov{\pi}.
\ee
To obtain variations of higher derivatives, differentiate each side of (\ref{emQED}). Then the second time derivatives will be expressed by the fields and their first derivatives, whose variations are already known. Variation of the resulting system leads to
\be
\delta\ddot{A}_0=\dot{u}^0,\quad \delta\ddot{A}_j=\partial_j(\delta u^0-\delta\dot{w}),\quad,
\delta\ddot{\psi}=-ie(\delta u^0-\delta\dot{w})\psi-e^2(\delta w)^2\psi+2ie\delta w\dot{\psi},\quad,
\delta\ddot{\pi}^j=0
\ee
(variations of the derivatives of the remaining fields can be found by conjugations and application of the constraints). Note that $\delta u^0$ and $\delta w$ are not infinitesimal and therefore some care is required in calculating the variations. One should simply calculate the difference, e.g. $\delta\ddot{\psi}=\ddot{\tilde{\psi}}-\ddot{\psi}$, so that the terms proportional to the higher powers of $\delta u^0$ and $\delta w$ are not omitted. 

The variations of higher derivatives can be found by continuation of this iteration procedure. The Tylor series that emerges can by shrunk to
\be\label{gauge}
\begin{aligned}
&\delta A_0(t,\vec{x})=\partial_t\lambda(t,\vec{x}),\qquad 
\delta A_j(t,\vec{x})=\partial_j\l{\lambda(t,\vec{x})+\kappa(t,\vec{x})}\r,\qquad
 \delta\psi(t,\vec{x})=\left({e^{-ie(\lambda(t,\vec{x})+\kappa(t,\vec{x}))}-1}\right)\psi(t,\vec{x}),\\
&\lambda(t,\vec{x}):=\int_{t_0}^t dt'\int_{t_0}^{t'} dt'' \delta u^0(t'',\vec{x}),\qquad
\kappa(t,\vec{x}):=-\int_{t_0}^t dt'\delta w(t',\vec{x}) .
\end{aligned}
\ee
The transformation of $\pi^j$ can be found from (\ref{emQED}b). It follows that $\delta\pi^j(t,\vec{x})=0$. Hence, it is $\pi^j=F^{j0}+\partial_j w$, and not $F^{j0}$, which is gauge invariant and hence can be interpreted as a measurable physical electric field $E^j$. If this interpretation is assumed than the equations of motion that follow from the extended formalism, when expressed in terms of measurable quantities $\vec{E}$, $\vec{B}$, $j^a=e\ov{\psi}\gamma^a\psi$ are precisely the Maxwell equations (note that from (\ref{gauge}) it follows that $j^a$ is indeed gauge invariant).

\section{Coulomb gauge}\label{Coul}

Now that we have all the gauge freedom explicitly described, we should proceed to quantize the theory. Since the only second class constraints are $\chi_1$ and $\chi_2$, the generalized Dirac bracket is given simply by
the formula (III.23) of \cite{Kazm6}, although the generalized Poisson bracket is now given by (\ref{GPBQED}).
If GDB's are promoted to the commutators of the operators in the quantum theory, the second class constraints can be consistently interpreted as strong operator equalities. Were there no first class constraints, we could try to quantize the theory in much the same way as we did for the free Dirac field in \cite{Kazm6}. But the first class constraints are the obstacle. They cannot be interpreted as strong operator equations, since their GDB's with other dynamical variables do not vanish in general. One method of implementation of these constraints in the quantum theory is to demand that  the {\it physical Hilbert space} of physical quantum states is a subset of a larger {\it kinematical Hilbert space} defined by the condition $\gamma_a|\psi\rangle =0$, where $a$ enumerates all the first class constraints. This method was originally proposed by Dirac \cite{Dirac2} and gave rise to the BRST quantization. However, I shall use a different method, which allows the computation of physically measurable quantities most quickly. This is the fixed gauge quantization. In this approach, one simply adds another constraint to the theory, called  {\it gauge condition}, which results in all the constraints being second class at the end. A gauge condition is any relation between the $q$'s and the $p$'s of the form 
\be\label{gengaugecon}
\chi(q,p)\approx 0
\ee
 which is accessible, i.e. any point $(q,p)$ can be transformed by gauge transformation to the one that satisfies (\ref{gengaugecon}). Also, we wish that the gauge condition, after being inserted to the consistency algorithm together with other constraints, eliminate all the gauge freedom. Otherwise, we need to impose farther gauge conditions or use the Dirac quantization anyway.

A shall consider the simplest condition, which is the Coulomb gauge
\be\label{coulomb}
\chi=\partial_i A^i\equiv \vec{\nabla} \vec{A}\approx 0.
\ee
This is clearily accessible by the transformation (\ref{gauge}), since for any $\vec{A}$ one can construct 
$\vec{A}'=\vec{A}-\vec{\nabla}f$, $f=\lambda+\kappa$ that satisfies $\vec{\nabla}\vec{A}=0$. One only needs to take
\be
f(t,\vec{x})=-\frac{1}{4\pi}\int\frac{\vec{\nabla}\vec{A}(t,\vec{x}')}{|\vec{x}-\vec{x}'|}d^3\vec{x}' .
\ee
To see this, use the identity
\be\label{identity}
-\triangle_{\vec{x}}\frac{1}{|\vec{x}-\vec{y}|}=
\vec{\nabla}_{\vec{x}}\l{\frac{\vec{x}-\vec{y}}{|\vec{x}-\vec{y}|^3}}\r=4\pi\delta^3(\vec{x}-\vec{y}).
\ee

Let us now step back to the point of the analysis in which the primary constraints were just found, eq. (\ref{primaryQED}). We shall now add the gauge condition (\ref{coulomb}) to the set of primary constraints (the reader is encouraged to find out that the results would not alter if $\chi$ was interpreted as a secondary constraint). Hence, the primary constraints and the total Hamiltonian are now
\be\label{primaryCoul}
\ba
&\chi_1=\pi-\frac{i}{2}\overline{\psi}\gamma^0,\quad \chi_2=\ov{\pi}+\frac{i}{2}\gamma^0\psi,\quad
\gamma_1=\pi^0,\quad \chi=\partial_jA^j, \\
&H_T=H+\int\l{u^0(\vec{x})\gamma_1(\vec{x})+\chi_1(\vec{x})u^1(\vec{x})+
u^2(\vec{x})\chi_2(\vec{x})+u(\vec{x})\chi(\vec{x})}\r, d^3x ,
\ea
\ee
where $H$ is still given by (\ref{H}). Now that the set of primary constraints and $H_T$ have changed, it is necessary to rerun the consistency algorithm for the time evolution of constraints. As before, the bracket of $\pi^0$ with $H_T$ gives the constraint $\tilde{\gamma}_2$ and requiring the brackets of $\chi_1$ and $\chi_2$ with $H_T$ to vanish leads to the restrictions on $u^1$ and $u^2$ (\ref{U1U2}). However, calculation of the bracket of $\chi$ with $H_T$ results in a new constraint
\be
[\chi,H_T]_{GP}=[\partial_j A^j,H_T]_{GP}=-\partial_j\frac{\delta H_T}{\delta \pi^j}=-\partial_j\pi^j-\triangle A_0=:\phi\approx 0.
\ee
So, in the first run of imposing consistency conditions the two secondary constraints $\tilde{\gamma}_2$ and $\phi$ were found and the restrictions (\ref{U1U2}) on $u^1$ and $u^2$. Continuing with the algorithm for $\tilde{\gamma}_2$ and $\phi$ one gets
\be
\begin{aligned}
&[\tilde{\gamma}_2,H_T]_{GP}=-e\left[{\partial_j\l{\ov{\psi}\gamma^j\psi}\r+\ov{\psi}\gamma^0u^1+u^2\gamma^0\psi}\right]
+\partial_j\partial_j u\approx 0 ,\\
&[\phi,H_T]_{GP}=e\partial_j\l{\ov{\psi}\gamma^j\psi}\r+\partial_j\partial_j (u-u^0).
\end{aligned}
\ee
If (\ref{U1U2}) is used (which is allowed), the first equation gives the restriction on $u$
\be
\triangle u=0 .
\ee
The second equation than gives
\be
\triangle u^0=e\partial_j\l{\ov{\psi}\gamma^j\psi}\r .
\ee
Hence, no farther constraints were produced and the consistency algorithm is now finished. Instead of $\tilde{\gamma_2}$ the constraint $\gamma_2=\partial_j\pi^j+ie(\pi\psi-\ov{\psi}\ov{\pi})$ can be used, as before. The commutation relations of the constraints are now the following: $\pi^0$ commutes with everything but $\phi$
\be
[\pi^0(\vec{x}),\phi(\vec{y})]_{GP}=\triangle\delta(\vec{x}-\vec{y}) .
\ee
$\gamma_2$ commutes with everything but $\chi$
\be
[\gamma_2(\vec{x}),\chi(\vec{y})]_{GP}=-\triangle\delta(\vec{x}-\vec{y}) .
\ee
$\chi_1$ and $\chi_2$ commute with all the other constraints and the bracket of $\chi_1$ with $\chi_2$ is the same as before. Finally, the bracket of $\chi$ with $\phi$ is
\be
[\chi(\vec{x}),\phi(\vec{y})]_{GP}=-\triangle\delta(\vec{x}-\vec{y}) .
\ee
Hence, all the constraints are now second class, as desired. It is straightforward to check that these constraints do not hide any first class ones, i.e. it is impossible to construct a first class constraint from a linear combination of them.

\subsection{The generalized Dirac bracket in the Coulomb gauge}

Before calculating the GDB, it is useful to rearrange slightly and rename the constraints. In the following I shall use
\be
\ba
&\chi_1=\pi-\frac{i}{2}\ov{\psi}\gamma^0,\qquad \chi_2=\ov{\pi}+\frac{i}{2}\gamma^0\psi,\\
&\chi_3\equiv \gamma_2=\partial_j\pi^j+ie(\pi\psi-\ov{\psi}\ov{\pi}),\qquad \chi_4\equiv\chi=\partial_jA^j,\\
&\chi_5\equiv\gamma_1=\pi^0,\qquad \chi_6=\phi+\gamma_2=ie(\pi\psi-\ov{\psi}\ov{\pi})-\triangle A_0.
\ea
\ee
Hence, instead of $\phi$ I will now use $\chi_6=\phi+\gamma_2$. This is because now the constraints can be grouped into pairs $\chi_1, \chi_2$; $\chi_3, \chi_4$; $\chi_5, \chi_6$;  such that the constraints in a given pair have vanishing brackets with those in the remaining pairs. The matrix $C_{\alpha\beta}$ will then acquire a block--diagonal form, which facilitates the calculation of its inverse. Explicitly, the non--vanishing brackets are
\be
\ba
&C_{1l\vec{x},2k\vec{y}}:=[\chi_{1l\vec{x}},\chi_{2k\vec{y}}]_{GP}=i\gamma^0_{kl}\delta(\vec{x}-\vec{y})=C_{2k\vec{x},1l\vec{y}},\\
&C_{3\vec{x},4\vec{y}}:=[\chi_{3\vec{x}},\chi_{4\vec{y}}]_{GP}=-\triangle\delta(\vec{x}-\vec{y})=-C_{4\vec{x},3\vec{y}}=-C_{4\vec{y},3\vec{x}},\\
&C_{5\vec{x},6\vec{y}}:=[\chi_{5\vec{x}},\chi_{6\vec{y}}]_{GP}=\triangle\delta(\vec{x}-\vec{y})=-C_{6\vec{x},5\vec{y}}=-C_{6\vec{y},5\vec{x}}.
\ea
\ee
The inverse matrices to these blocks are then
\be\label{blockinv}
\ba
&C^{1l\vec{x},2k\vec{y}}=C^{2k\vec{y},1l\vec{x}}=-i\gamma^0_{lk}\delta(\vec{x}-\vec{y}),\\
&C^{3\vec{x},4\vec{y}}=-\frac{1}{4\pi|\vec{x}-\vec{y}|}+a_{34},\quad 
C^{4\vec{x},3\vec{y}}=\frac{1}{4\pi|\vec{x}-\vec{y}|}+a_{43},\\
&C^{5\vec{x},6\vec{y}}=\frac{1}{4\pi|\vec{x}-\vec{y}|}+a_{56},\quad 
C^{6\vec{x},5\vec{y}}=-\frac{1}{4\pi|\vec{x}-\vec{y}|}+a_{65} .
\ea
\ee
Here $a_{ij}$ are completely arbitrary numbers. It may be surprising that the inverse matrices are not determined in a unique way and that they even do not inherit the symmetries of the matrices to which they are inverse. This is because these are infinite--dimensional matrices to which standard theorems of linear algebra do not apply. That the results (\ref{blockinv}) are correct for arbitrary values of the constants $a_{ij}$ can be verified by direct computations, e.g.
\be
\int d^3y C^{3\vec{x},4\vec{y}}C_{4\vec{y},3\vec{z}}=
-\int d^3y\frac{1}{4\pi|\vec{x}-\vec{y}|}\triangle\delta(\vec{z}-\vec{y})+a_{34}\int d^3y\triangle\delta(\vec{z}-\vec{y})=-\triangle_{\vec{z}}\l{\frac{1}{4\pi|\vec{x}-\vec{z}}}\r=\delta(\vec{x}-\vec{z}).
\ee
That the term proportional to $a_{34}$ vanishes follows from the following reasoning: imagine that the area of integration is a bounded open subset of $\Omega\subset\mathbb{R}^3$ such that $\vec{z}\in\Omega$. Then the application of Gauss theorem (which is assumed to hold for distributions) allows to rewrite the integral of $\triangle\delta(\vec{z}-\vec{y})$ as a flux of the gradient $\vec{\nabla}_{\vec{y}}\delta(\vec{z}-\vec{y})$ through the boundary $\partial\Omega$. But the Dirac delta and its derivatives vanish everywhere except the points in which their argument is zero. Hence, the integration of $\vec{\nabla}_{\vec{y}}\delta(\vec{z}-\vec{y})$ with respect to $\vec{y}$ over the region that does not contain $\vec{z}$ (such as $\partial\Omega$) must necessarily give zero. Now for any $\vec{z}\in\mathbb{R}^3$ one can find $\Omega\ni\vec{z}$ and decompose the integral over $\mathbb{R}^3$ into the one over $\Omega$ and $\mathbb{R}^3\setminus\Omega$. The integral over the complement of $\Omega$ vanishes, since there the argument of $\triangle\delta$ is always nonzero. This completes the argument.

So, the arbitrary constants appeared in the inverse matrix $C^{\alpha\beta}$. Will then the GDB also include them? The answer is no! Recall that the the purpose for the particular construction of GDB was to enable consistent replacement of the brackets by the commutators or anti--commutators of operators, dependently on whether the variables are even or odd. In order for this to be possible, the GDB was required to posses appropriate symmetries. This symmetries will not be satisfied if the inverse matrix (\ref{blockinv}) is used in the construction, unless
\be
a_{33}=a_{44}=a_{55}=a_{66}=0,\qquad a_{43}=-a_{34},\qquad a_{65}=-a_{56} .
\ee
So the freedom in the construction of the GDB is now reduced to the two parameters $a_{34}$ and $a_{56}$. The partially solved formula for the GDB that is convenient for performing farther calculations is then
\be\label{GDBcoul}
\ba
&[F,G]_{GD}=[F,G]_{GP}+i\gamma_{kk'}\int d^3x 
\l{[F,\chi_{1k\vec{x}}]_{GP}[\chi_{2k'\vec{x}},G]_{GP}+[F,\chi_{2k'\vec{x}}]_{GP}[\chi_{1k\vec{x}},G]_{GP}}\r\\
&+\iint d^3xd^3y
\l{
\frac{1}{4\pi|\vec{x}-\vec{y}|}-a_{34}
}\r
\l{
\frac{\partial}{\partial y^k}\frac{\delta F}{\delta\pi^k(\vec{y})}[G,\gamma_2(\vec{x})]_{GP}-[F,\gamma_2(\vec{x})]_{GP}\frac{\partial}{\partial y^k}\frac{\delta G}{\delta\pi^k(\vec{y})}
}\r \\
&+\iint d^3xd^3y
\l{
\frac{1}{4\pi|\vec{x}-\vec{y}|}+a_{56}
}\r
\l{
\frac{\delta F}{\delta A_0(\vec{x})}[G,\chi_6(\vec{y})]_{GP}-[F,\chi_6(\vec{y})]_{GP}\frac{\delta G}{\delta A_0(\vec{x})}
}\r
 .
\ea
\ee
However, another requirement for the GDB was that it is consistent with all the second class constraints. It appears that there is a problem with (\ref{GDBcoul}) with this respect. To see this, take $F=A_0(\vec{z})$ and $G=\int d^3x\lambda(\vec{x})\psi(\vec{x})$, where $\lambda$ is arbitrary function that does not depend on the canonical fields. It follows from (\ref{GDBcoul}) that
\be\label{A0G1}
[A_0(\vec{z}),G]_{GD}=ie\int d^3x\l{\frac{1}{4\pi|\vec{x}-\vec{z}|}+a_{56}}\r\lambda(\vec{x})\psi(\vec{x}) .
\ee
But the constraint $\chi_6$ tells us that $\triangle A_0=ie(\pi\psi-\ov{\psi}\ov{\pi})$, which, under the physically reasonable assumption that $A_0$ is bounded everywhere, can be integrated to yield
\be\label{A0}
A_0(\vec{z})=\frac{e}{4\pi}\int\frac{\ov{\psi}(\vec{z}')\gamma^0\psi(\vec{z}')}{|\vec{z}-\vec{z}'|}d^3z'+a_{\infty} ,
\ee
where the constant $a_{\infty}$ does not contribute anything to the brackets and fact could be set to zero without any loss of generality in the subsequent analysis. If (\ref{A0}) is used, the calculation of the bracket now gives
\be\label{A0G2}
[A_0(\vec{z}),G]_{GD}=ie\int\frac{ d^3x}{4\pi|\vec{x}-\vec{z}|}\lambda(\vec{x})\psi(\vec{x}) .
\ee
Clearly, (\ref{A0G1}) agrees with  (\ref{A0G2}) if and only if $a_{56}=0$! This is a strange result, since the bracket (\ref{GDBcoul}) of any dynamical variable with any of the constraints $\chi_i$, $i=1,2,3,4,5,6$ can be verified to vanish for ARBITRARY values of $a_{34}$ and $a_{56}$. So it seams that it is consistent with the constraints! The point is that in obtaining (\ref{A0}) I did not use the constraint $\chi_6$ but its integrated version
\be
\tilde{\chi}_{6}(\vec{y}):=\int\frac{\chi_6(\vec{x}) d^3x}{4\pi|\vec{x}-\vec{y}|}=ie\int \frac{\l{\pi\psi-\ov{\psi}\ov{\pi}}\r(\vec{x})}{4\pi|\vec{x}-\vec{y}|}d^3x+A_0(\vec{y})-a_{\infty} .
\ee
The bracket (\ref{GDBcoul}) of a dynamical variable with this integrated constraint does not vanish in general, unless $a_{56}=0$. But how it can be that the bracket with $\chi_6$ vanishes but the bracket with the integral of $\chi_6$ does not? Does this fact contradict the linearity of the bracket? This apparent paradox can be traced back to the following calculation (certainly incorrect)
\be\label{paradox}
0=\int\frac{d^3y}{4\pi|\vec{z}-\vec{y}|}\int d^3x\triangle\delta(\vec{x}-\vec{y})=\int d^3x\int d^3y\frac{\triangle\delta(\vec{x}-\vec{y})}{4\pi|\vec{z}-\vec{y}|} .
\ee\label{pomparad}
The first equality stems from the fact that for any $\vec{y}\in\mathbb{R}^3$ we have  $\int d^3x\triangle\delta(\vec{x}-\vec{y})=0$ (use Gauss theorem). That the result is wrong can be seen by substituting
\be\frac{\triangle\delta(\vec{x}-\vec{y})}{4\pi|\vec{z}-\vec{y}|}=
\frac{\partial}{\partial y^i}\left[{\frac{1}{4\pi|\vec{z}-\vec{y}|}\frac{\partial}{\partial y^i}\delta(\vec{y}-\vec{x})-\frac{\partial}{\partial y^i}\l{\frac{1}{4\pi|\vec{z}-\vec{y}|}}\r\delta(\vec{y}-\vec{x})}\right]
+\frac{\partial}{\partial y^i}\frac{\partial}{\partial y^i}\l{\frac{1}{4\pi|\vec{z}-\vec{y}|}}\r \delta(\vec{y}-\vec{x})
\ee
into the RHS of (\ref{paradox}). Then the integral of the first component of (\ref{pomparad}) vanishes, since this component is the divergence of a vector field that vanishes everywhere except the point $\vec{y}=\vec{x}$. The integral of the second component of (\ref{pomparad}) gives $-1$, since this component is just equal to $-\delta(\vec{z}-\vec{y})\delta(\vec{y}-\vec{x})$, on account of (\ref{identity}). Hence, we finally obtain from (\ref{paradox}) that $0=-1$, which seems incorrect. The mistake was made in the second equality of (\ref{paradox}), where I tacitly assumed that the order of integration with respect to $\vec{x}$ and $\vec{y}$ can be swapped. It appears that it can not. This is the reason why the linearity of the brackets that involve integrations in their structure cannot be naively exploited.

In order to be able to use both the differential and integrated form of the constraints consistently with the bracket, I will set $a_{34}=a_{56}=0$. Finally then the GDB does not contain any arbitrariness and is given by
\be\label{GDBost}
\ba
&[F,G]_{GD}=[F,G]_{GP}+i\gamma^0_{kk'}\int d^3x 
\l{[F,\chi_{1k\vec{x}}]_{GP}[\chi_{2k'\vec{x}},G]_{GP}+[F,\chi_{2k'\vec{x}}]_{GP}[\chi_{1k\vec{x}},G]_{GP}}\r +\\
&\iint 
\frac{d^3xd^3y}{4\pi|\vec{x}-\vec{y}|}
\left({
\frac{\partial}{\partial y^k}\frac{\delta F}{\delta\pi^k(\vec{y})}[G,\gamma_2(\vec{x})]_{GP}-[F,\gamma_2(\vec{x})]_{GP}\frac{\partial}{\partial y^k}\frac{\delta G}{\delta\pi^k(\vec{y})}
+
\frac{\delta F}{\delta A_0(\vec{x})}[G,\chi_6(\vec{y})]_{GP}-[F,\chi_6(\vec{y})]_{GP}\frac{\delta G}{\delta A_0(\vec{x})}
}\right) 
\ea
\ee
or, even more explicitly,
\be\label{GDBost2}
\ba
&[F,G]_{GD}=[F,G]_{GP}+i\gamma^0_{kk'}\int d^3x 
\l{[F,\chi_{1k\vec{x}}]_{GP}[\chi_{2k'\vec{x}},G]_{GP}+[F,\chi_{2k'\vec{x}}]_{GP}[\chi_{1k\vec{x}},G]_{GP}}\r +
\iint 
\frac{d^3xd^3y}{4\pi|\vec{x}-\vec{y}|}\\
&\left[{
\frac{\partial}{\partial y^k}\frac{\delta F}{\delta\pi^k(\vec{y})} \ 
\frac{\partial}{\partial x^j}\frac{\delta G}{\delta A_j(\vec{x})}-
\frac{\partial}{\partial x^j}\frac{\delta F}{\delta A_j(\vec{x})} \ 
\frac{\partial}{\partial y^k}\frac{\delta G}{\delta\pi^k(\vec{y})}
+
\frac{\delta F}{\delta A_0(\vec{y})}\triangle_{\vec{x}}\l{\frac{\delta G}{\delta \pi^0(\vec{x})}}\r-
\triangle_{\vec{x}}\l{\frac{\delta F}{\delta \pi^0(\vec{x})}}\r \frac{\delta G}{\delta A_0(\vec{y})}
}\right.\\
&\left.{
+
ie\l{\frac{\partial}{\partial y^k}\frac{\delta F}{\delta\pi^k(\vec{y})}+\frac{\delta F}{\delta A_0(\vec{y})}}\r
[G,(\pi\psi-\ov{\psi}\ov{\pi})(\vec{x})]_{GP}-
ie[F,(\pi\psi-\ov{\psi}\ov{\pi})(\vec{x})]_{GP}
\l{\frac{\partial}{\partial y^k}\frac{\delta G}{\delta\pi^k(\vec{y})}+\frac{\delta G}{\delta A_0(\vec{y})}}\r
}\right]
 .
\ea
\ee
From now on, the canonical variables $A_a$, $\pi^a$ will be referred to as electromagnetic variables, whereas $\psi$, $\ov{\psi}$, $\pi$ and $\ov{\pi}$ as spinor variables. \newline
1) If both $F$ and $G$ depend on spinor variables only, then the only contribution to the bracket comes from the first line. The bracket is the same as the one calculated in the free Dirac field theory considered in \cite{Kazm6}. \newline
2) Consider electromagnetic variables. It is clear that from (\ref{GDBost2}) that $[A_a(\vec{x}),A_b(\vec{y})]_{GD}=[\pi^a(\vec{x}),\pi^b(\vec{y})]_{GD}=0$. The bracket of $\pi^0$ with any other variable is certainly zero, since $\pi^0$ is a constraint. 
The bracket of $A_i$ with any variable $G$ is 
\be\label{AG}
[A_i(\vec{x}),G]_{GD}=\frac{\delta G}{\delta \pi^i(\vec{x})}+\int d^3y
\frac{\partial}{\partial x^i}\l{\frac{1}{4\pi|\vec{x}-\vec{y}|}}\r
\frac{\partial}{\partial y^k}\l{\frac{\delta G}{\delta\pi^k(\vec{y})}}\r ,
\ee
from which the bracket for the canonically conjugate pair of electromagnetic variables follows
\be
[A_i(\vec{x}),\pi^j(\vec{y})]_{GD}=
\delta^j_i\delta(\vec{x}-\vec{y})+\frac{\partial}{\partial x^i}\frac{\partial}{\partial x^j}
\l{\frac{1}{4\pi |\vec{x}-\vec{y}|}}\r .
\ee
The bracket of $A_0$ with $G_{EM}$ that depends on the electromagnetic variables only is
\be
[A_0(\vec{y}),G_{EM}]_{GD}=\int d^3x \frac{\partial}{\partial x^j}
\left[{
\frac{1}{4\pi|\vec{x}-\vec{z}|}\frac{\partial}{\partial x^j}\l{\frac{\delta G_{EM}}{\delta\pi^0(\vec{x})}}\r
-\frac{\partial}{\partial x^j}\l{\frac{1}{4\pi|\vec{x}-\vec{z}|}}\r\frac{\delta G_{EM}}{\delta\pi^0(\vec{x})}
}\right] .
\ee
The integrand is a divergence of a vector field and hence the expression vanishes if $\delta G_{EM}/\delta\pi^0(\vec{x})$ has bounded support, or just tends to zero sufficiently fast with $\vec{x}$ going to infinity. Note however that if, say, $G_{EM}=\int\lambda(\vec{x})\pi^0(\vec{x})d^3x$, where $\lambda$ does not tend to zero at infinity, then the bracket of $G_{EM}$ with $A_0(\vec{y})$ will not vanish! So, the smeared constraints are consistent with the bracket (\ref{GDBost2}) if and only if the smearing functions decrease sufficiently rapidly at infinity.\newline
3) The bracket of $A_0$ with $G_{SP}$ that depends on the spinor variables only is
\be
[A_0(\vec{y}),G_{SP}]_{GD}=ie\int\frac{d^3x}{4\pi|\vec{x}-\vec{y}|}[G_{SP},(\pi\psi-\ov{\psi}\ov{\pi})(\vec{x})]_{GP} .
\ee
4) The bracket of $F_{SP}$ that depends on the spinor variables only with $\pi^j(\vec{y})$ is
\be
[F_{SP},\pi^j(\vec{y})]_{GD}=ie\int d^3x\frac{\partial}{\partial y^j}\l{\frac{1}{4\pi|\vec{x}-\vec{y}|}}\r
[F_{SP},\pi(\vec{x})\psi(\vec{x})-\ov{\psi}(\vec{x})\ov{\pi}(\vec{x})]_{GP} ,
\ee
which is the same as $-[F_{SP},\partial_jA_0(\vec{y})]_{GD}$. This observation led Weinberg \cite{Wein} to define a combined variable
\be\label{piperp}
\pi_{\perp}^{j}:=\pi^j+\partial_jA_0,
\ee
which has trivial bracket with spinor variables
\be
[F_{SP},\pi_{\perp}^{j}(\vec{y})]_{GD}=0 .
\ee
Note that from $\chi_3=0$ and $\chi_6=0$ it follows that
\be
\partial_j\pi_{\perp}^j=0 .
\ee
The bracket of $\pi_{\perp}^{j}$ with arbitrary dynamical variable $F$ is
\be\label{Fpiperp}
[F,\pi_{\perp}^{j}(\vec{y})]_{GD}=\frac{\delta F}{\delta A_j(\vec{y})}-\int d^3x
\l{\frac{\partial}{\partial x^k}\frac{\delta F}{\delta A_k(\vec{x})}}\r
\l{\frac{\partial}{\partial x^j}\frac{1}{4\pi|\vec{x}-\vec{y}|}}\r ,
\ee
from which it follows that
\be\label{Api}
[A_i(\vec{x}),\pi_{\perp}^j(\vec{y})]_{GD}=
\delta^j_i\delta(\vec{x}-\vec{y})+\frac{\partial}{\partial x^i}\frac{\partial}{\partial x^j}
\l{\frac{1}{4\pi |\vec{x}-\vec{y}|}}\r .
\ee

\subsection{The equations of motion}
Since the GDB is now used that is consistent with the second class constraints and all the constraints are now second class, we can freely use them to simplify the form of observables that are of interest. For example, the distinction between canonical, first class, total and extended Hamiltonians is now spurious: the simplest form of the Hamiltonian can be given as
\be\label{Hcoul}
\begin{aligned}
&H=\int \l{\mathcal{H}_{EM}(\vec{x})+\mathcal{H}_D(\vec{x})+\mathcal{H}_{I}(\vec{x})}\r d^3x,\\
&\mathcal{H}_{EM}=\frac{1}{2}\pi_{\perp}^{j}\pi_{\perp}^{j}+\frac{1}{4}F_{ij}F_{ij},\qquad
\mathcal{H}_D=\ov{\psi}\l{-i\gamma^j\partial_j+m}\r\psi,\qquad
\mathcal{H}_{I}=eA_j\ov{\psi}\gamma^j\psi+\frac{1}{2}eA_0\ov{\psi}\gamma^0\psi .
\end{aligned}
\ee
The equations of motion for the canonical variables are
\be\label{EMcoul}
\begin{aligned}
&\dot{A}_0(\vec{y})=[A_0(\vec{y}),H]_{GD}=-e\int\frac{\partial_j(\ov{\psi}\gamma^j\psi)(\vec{x})}{4\pi|\vec{x}-\vec{y}|}d^3x, \\
&\dot{A}_i(\vec{y})=[A_i(\vec{y}),H]_{GD}=\pi_{\perp}^i(\vec{y}),\\
&\dot{\pi}^i(\vec{y})=[\pi^i(\vec{y}),H]_{GD}=\partial_jF_{ji}(\vec{y})-e\l{\ov{\psi}\gamma^i\psi}\r(\vec{y}) 
\quad \Longrightarrow \  \dot{\vec{E}}=\vec{\nabla}\times \vec{B}-\vec{J}\\
&\dot{\psi}(\vec{y})=-\gamma^0\gamma^j\mathcal{D}_j\psi(\vec{y})-ieA_0(\vec{y})\psi(\vec{y})-im\gamma^0\psi(\vec{y}) \ \Longrightarrow \ 
\l{i\gamma^a\mathcal{D}_a-m}\r\psi=0
\end{aligned}
\ee
The equations certainly need to be supplemented by the constraints. The constraints $\chi_1$ and $\chi_2$ can be used just to eliminate the variables $\pi$ and $\ov{\pi}$ from the formalism once and for all. Similarly, $\chi_5$ tells that $\pi^0=0$ is not a physical degree of freedom. The constraint $\chi_3$ then reduces to the Gauss law $\vec{\nabla}\vec{\pi}=e\ov{\psi}\gamma^0\psi$ and $\chi_6$ tells that $A_0$ is not an independent variable but a functional of matter fields
\be\label{A0ost}
A_0(\vec{x})=e\int\frac{\l{\ov{\psi}\gamma^0\psi}\r(\vec{y})}{4\pi|\vec{x}-\vec{y}|}d^3y .
\ee
For simplicity, I imposed on $A_0$ the condition of vanishing at infinity. The general bounded solution to the constraint is obviously (\ref{A0}). The reader is encouraged to verify the effects produced by nonzero constant $a_{\infty}$ of (\ref{A0}) in the final results. 

Note farther that (\ref{piperp}) allows for the elimination of $\pi^j$ in favor of $\pi_{\perp}^j$ and $\psi$, $\ov{\psi}$. Then $\pi_{\perp}^j$ can be eliminated in favor of $\dot{A}_j$ on account of the second equation of (\ref{EMcoul}). Finally, the only electromagnetic physically important fields are $A_j$'s, which are restricted by the Coulomb gauge condition $\chi_4=\partial_jA^j=0$, so we end up with the two degrees of freedom of the electromagnetic field, as desired.

\subsection{Transition to the interaction picture}

In order to construct the quantum theory in the case of the free Dirac field \cite{Kazm6}, we had to find a general solution to the evolution equations for fields. The arbitrary operator coefficients in the general solution appeared to obey an exceptionally simple commutation rules with themselves and the Hamiltonian and other physically important observables such as momentum or electric charge operator (although these commutation relations were not explicitly verified in \cite{Kazm6}). This simplicity was crucial and allowed the construction of Fock space carrying the representation of all the commutation relations following from the GDB.

Even the first look on (\ref{EMcoul}) allows to see that no such simple solution to the system of equations for QED in the Coulomb gauge is possible. One could try to differentiate the second equation and then use the third in order to eliminate $\pi^j$ from the system. Also, the first equation can be used to eliminate $A_0$. This can be done, but the resulting relation between $A_j$ and $\psi$, $\ov{\psi}$ is fairly too complicated to be exactly solvable.

The way out of this difficulties is to pass to the interaction picture and calculate physical quantities perturbatively. To accomplish this, let us decompose the Hamiltonian (\ref{Hcoul}) into the free part $H_0$ and the interaction part $H_I$ according to 
\be\label{H0HI}
\ba
H_0(t)=\int (\mathcal{H}_{EM}(t,\vec{x})+\mathcal{H}_D(t,\vec{x})) d^3 x,\qquad
H_I(t)=\int\mathcal{H}_I(t,\vec{x}) d^3 x,\qquad
H=H_0(t)+H_I(t) .
\ea
\ee
Up to this point, I did not even bother to write the time argument of fields. It was understood that $\psi(\vec{x})$, $A_j(\vec{x})$, etc. depend on time and their evolution is determined by the total Hamiltonian. Now the Hamiltonian itself is time--independent, but its constituent parts, such as $H_0$ and $H_I$ do depend on time, as indicated in (\ref{H0HI}). I shall now chose an instant of time, for simplicity let it be $t=0$. From now on $H_0$ and $H_I$ should be understood as $H_0(0)$ and $H_I(0)$ (the same concerns $H_D$ and $H_{EM}$). Let $F(t)$ be a dynamical variable, whose time evolution is determined in a usual way, by $H$. By the {\it interaction picture} of $F$ we will understand a variable $F_{(I)}(t)$ whose evolution is determined by $H_0$ and whose value at $t=0$ is equal to the value of $F$ at this instant:
\be
\ba
&F_{(I)}(0)=F(0),\\
&\dot{F}_{(I)}(0)=\left[{F_{(I)}(0),H_0}\right]_{GD},\\
&\ddot{F}_{(I)}(0)=\left[{\dot{F}_{(I)}(0),H_0}\right]_{GD}=\left[{\left[{F_{(I)}(0),H_0}\right]_{GD},H_0}\right]_{GD},\quad  \\
&\vdots \\
&{F}^{(n)}_{(I)}(0)=\left[{F^{(n-1)}_{(I)}(0),H_0}\right]_{GD}=\left[{\cdots\left[{\left[{F_{(I)}(0),H_0}\right]_{GD},H_0}\right]_{GD}\cdots H_0}\right]_{GD} .
\ea
\ee
From this definition it follows that
\be
\dot{F}_{(I)}(t)=\left[{F_{(I)}(t),H_0}\right]_{GD}
\ee
for any $t$ (use the Tylor expansion). Another immediate consequences of the definition are that the time derivative of an interaction picture variable also evolves in the interaction picture (the same is obviously true for spatial partial derivatives). Finally, for any variables $F$ and $G$ the following implication holds
\be\label{impliInt}
\dot{F}_{(I)}(0)=\left[{F_{(I)}(0),H_0}\right]_{GD}=G(0) \ \Longrightarrow \ \dot{F}_{(I)}(t)=G_{(I)}(t)
\ee
for any $t$. This latter corollary will be used repeatably below (the relations between the derivatives at $t=0$ will be obtained and the relations between the interaction picture fields at any time will then be assumed).

Using (\ref{Hcoul}) and (\ref{GDBost2}), the GDB of any dynamical variable $F$ with $H_0$ can be calculated
\be\label{FH0}
\ba
&[F,H_0]_{GD}=[F,H_0]_{GP}+
i\gamma^0_{ll'}\int\l{
[F,\chi_{1l\vec{x}}]_{GP}[\chi_{2l'\vec{x}},H_D]_{GP}+[F,\chi_{2l'\vec{x}}]_{GP}[\chi_{1l\vec{x}},H_D]_{GP}
}\r d^3x \\
&+
ie\iint\frac{d^3xd^3y}{4\pi|\vec{x}-\vec{y}|}
\l{
\frac{\partial}{\partial y^k}\frac{\delta F}{\delta \pi^k(\vec{y})}+\frac{\delta F}{\delta A_0(\vec{y})}
}\r
[H_D,(\pi\psi-\ov{\psi}\ov{\pi})(\vec{x})]_{GP} .
\ea
\ee
This simplification follows from 
\be
\frac{\delta H_0}{\delta\pi^k}=\pi_{\perp}^k \Rightarrow \partial_k\frac{\delta H_0}{\delta\pi^k}=0,\qquad
\frac{\delta H_0}{\delta A_0}=-\partial_j\pi_{\perp}^j=0,\qquad
\frac{\delta H_0}{\delta A_j}=-\partial_iF_{ij} \Rightarrow \partial_j\frac{\delta H_0}{\delta A_j}=0 .
\ee
From (\ref{FH0}) it follows that if $F_{SP}$ depends on the spinor variables only then its bracket is the same as the one that would be obtained in the theory of the free Dirac field discussed in \cite{Kazm6}. Hence, the interaction picture field $\psi_{(I)}$ satisfies the free Dirac equation
\be\label{dirI}
\l{i\gamma^a\partial_a-m}\r\psi_{(I)}=0 .
\ee
For other fields, the calculations give
\be\label{emIP}
\ba
&[A_i(0,\vec{x}),H_0]_{GD}=\pi_{\perp}^i(0,\vec{x}) \ \Longrightarrow \ \dot{A}_{(I)i}(t,\vec{x})=\pi_{\perp(I)}^i(t,\vec{x}), \\
&[\pi_{\perp}^i(0,\vec{x}),H_0]_{GD}=\partial_jF_{ji}(0,\vec{x}) \ \Longrightarrow \ \dot{\pi}_{\perp(I)}^i(t,\vec{x})=\partial_jF_{(I)ji}(t,\vec{x}) ,\\
&[A_0(0,\vec{x}),H]_{GD}=-e\int\frac{\partial_j(\ov{\psi}\gamma^j\psi)(0,\vec{y})}{4\pi|\vec{x}-\vec{y}|}d^3y
\ \Longrightarrow \ \dot{A}_{(I)0}(t,\vec{x})=-e\int\frac{\partial_j(\ov{\psi}_{(I)}\gamma^j\psi_{(I)})(t,\vec{y})}{4\pi|\vec{x}-\vec{y}|}d^3y .
\ea
\ee
The implication (\ref{impliInt}) was used. Note that the evolution of $A_0$ in the interaction picture is nontrivial, contrary to what seems to have been suggested in \cite{Wein}. However, the point is that $A_0$ is no longer necessary, since the first and the second equations now give a simple equation for $A_{(I)i}$
\be\label{A1}
\ddot{A}_{(I)i}=\partial_jF_{(I)ji}=\triangle A_{(I)i} \ \Longrightarrow \ \square A_{(I)i}=0 .
\ee
This equation, together with the Coulomb gauge constraint
\be\label{A2}
\partial_jA_{(I)j}=0
\ee
and the equation (\ref{dirI}) for $\psi_{(I)}$ are the only important equations for the evolution of interaction picture fields. Now these equations are sufficiently simple that the general solution can be readily found. Indeed, (\ref{dirI}) have already been solved in \cite{Kazm6}.  The general solution to (\ref{A1}) and (\ref{A2}) is
\be\label{AI}
A_{(I)i}(x)=\int d\Gamma_k\l{e^{-ik\cdot x}e_{i\lambda}(\vec{k})b_{\lambda}(\vec{k})+e^{ik\cdot x}e^*_{i\lambda}(\vec{k})b^{\dag}_{\lambda}(\vec{k})}\r ,\qquad k^ie_{i\lambda}(\vec{k})=0 .
\ee
The assumption is made that $k^0=|\vec{k}|$ and $d\Gamma_k=d^3k/2\pi^32k^0$, in complete parallel with the conventions of \cite{Kazm6}, which were adapted to the case of massless particles. Hence, $e_{\lambda}(\vec{k})$ for $\lambda=1,-1$ form a basis of the two--dimensional subspace of $\mathbb{R}^3$ orthogonal to $\vec{k}$ and $b_{\lambda}(\vec{k})$ are arbitrary coefficients. The coefficients $b^{\dag}_{\lambda}(\vec{k})$ are classically the complex conjugates of $b_{\lambda}(\vec{k})$, but the conjugation is denoted by $\dag$, since it will pass to Hermitian conjugation in the quantum theory, so that the operators $A_{(I)i}$ are self--adjoint. The letter $b$ was used to denote the coefficients in order to distinguish them from the creation and annihilation operators for the Dirac field that were introduced already in \cite{Kazm6}. A convenient choice of basis $e_{\lambda}$ is given by
\be\label{convbasis}
e_{\pm 1}(\vec{k})=\frac{1}{\sqrt{2}}
\l{
\begin{array}{ccc}
\cos \theta \cos \phi \mp i\sin \phi \\
\cos\theta\sin\phi\pm i\cos\phi \\
-\sin\theta
\end{array}
}\r ,\qquad
\vec{k}=|\vec{k}|
\l{
\begin{array}{ccc}
\sin \theta \cos \phi \\
\sin\theta\sin\phi \\
\cos\theta
\end{array}
}\r .
\ee
It is also convenient to define the matrices
\be
e(\vec{k}):=\l{e_1(\vec{k}),e_2(\vec{k})}\r=\frac{1}{\sqrt{2}}
\l{
\begin{array}{ccc}
\cos \theta \cos \phi - i\sin \phi \ & \ \cos \theta \cos \phi + i\sin \phi\\
\cos\theta\sin\phi + i\cos\phi \ & \ \cos\theta\sin\phi - i\cos\phi\\
-\sin\theta \ & \ -\sin\theta
\end{array}
}\r,\qquad
e_0:=\frac{1}{\sqrt{2}}
\l{
\begin{array}{ccc}
1 & 1\\
i & -i\\
0 & 0
\end{array}
}\r,
\ee
which are related by the standard rotation
\be
e(\vec{k})=R(\vec{k})e_0,\qquad
R(\vec{k}):=
\l{
\begin{array}{ccc}
\cos\phi & -\sin\phi & 0\\
\sin\phi & \cos\phi & 0\\
0 & 0 & 1
\end{array}
}\r
\l{
\begin{array}{ccc}
\cos\theta & 0 & \sin\theta\\
0 & 1 & 0\\
-\sin\theta & 0 & \cos\theta
\end{array}
}\r .
\ee
When these conventions are adopted, the relation (\ref{AI}) can be explicitly inverted with respect to the coefficients $b$, $b^{\dag}$ 
\be\label{binv}
\ba
&b(\vec{k})=\int d^3xe^{ik\cdot x}e_0^{\dag}R^{-1}(\vec{k})\l{k^0A_{(I)}(x)+i\dot{A}_{(I)}(x)}\r ,\\
&b^{\dag}(\vec{k})=\int d^3xe^{-ik\cdot x}e_0^{\dag}R^{-1}(\vec{k})\l{k^0A_{(I)}(x)-i\dot{A}_{(I)}(x)}\r .
\ea
\ee
Think of $b(\vec{k})$ as a column vector with components $b_{\lambda}(\vec{k})$ for $\lambda=1,-1$. Similarly, $A_{(I)}$ should be thought of as a column with components $A_{(I)1}, A_{(I)2}, A_{(I)3}$.

\subsection{Quantization}

\be
[\widehat{F},\widehat{G}]_{\mp}=i\widehat{[F,G]}_{GD} ,
\ee 
where the upper sign applies whenever at least one of the variables is even and the lower corresponds to the case of both variables being odd. The evolution of the Heisenberg and the interaction picture operators can now be written simply as
\be
\widehat{F}(t)=e^{it\widehat{H}}F(0)e^{-it\widehat{H}},\qquad
\widehat{F}_{(I)}(t)=e^{it\widehat{H}_0}\widehat{F}(0)e^{-it\widehat{H}_0}.
\ee
It follows that knowing the commutation relation at $t=0$ it is easy to obtain the one that is relevant for $t\not=0$ using
\be
[\widehat{F}(t),\widehat{G}(t)]_{\mp}=e^{it\widehat{H}}[\widehat{F}(0),\widehat{G}(0)]_{\mp}e^{-it\widehat{H}},\qquad
[\widehat{F}_{(I)}(t),\widehat{G}_{(I)}(t)]_{\mp}=e^{it\widehat{H}_0}[\widehat{F}_{(I)}(0),\widehat{G}_{(I)}(0)]_{\mp}e^{-it\widehat{H}_0}.
\ee
Hence, the relation (\ref{Api}) leads to
\be\label{ApiOp}
[\widehat{A}_{(I)i}(t,\vec{x}),\widehat{\pi}_{\perp(I)}^j(t,\vec{y})]_{-}=
[\widehat{A}_i(t,\vec{x}),\widehat{\pi}_{\perp}^j(t,\vec{y})]_{-}=
i\delta^j_i\delta(\vec{x}-\vec{y})+i\frac{\partial}{\partial x^i}\frac{\partial}{\partial x^j}
\l{\frac{1}{4\pi |\vec{x}-\vec{y}|}}\r .
\ee
Since $\dot{\widehat{A}}_{(I)i}(t,\vec{x})=\widehat{\pi}_{\perp(I)}^i(t,\vec{x})$ on account of (\ref{emIP}), the relation (\ref{ApiOp}), together with (\ref{binv}), can be used to derive the commutation relations of the operators corresponding to the coefficients $b$ and $b^{\dag}$ defined by (\ref{AI}) and (\ref{convbasis})
\be\label{comb}
[b_{\lambda}(\vec{k}),b_{\lambda'}(\vec{k}')]_{-}=\l{2\pi}\r^3 2k^0\delta_{\lambda\lambda'}\delta(\vec{k}-\vec{k}') 
\ee
(note that I do not use hats $\widehat{b}$ above the annihilation and creation operators).
The physical interpretation of $b$ and $b^{\dag}$ can be inspected in exactly the same way as the meaning of $a$ and $a^{\dag}$ was established in \cite{Kazm6}. Namely, the energy--momentum and spin density tensors should be constructed and their commutation relations with $b$ and $b^{\dag}$ determined. It follows that $\vec{k}$ ought to be interpreted as the momentum and $\lambda$ as the helicity of the particle created by $b_{\lambda}^{\dag}(\vec{k})$. Note, however, that this interpretation is valid in the asymptotic regions in which the interaction is sufficiently week that the time evolution can be satisfactorily approximated by the free Hamiltonian $\widehat{H}_0$. This situation is perfectly relevant for the description of scattering experiments.
Similarly, the operators $a_{\sigma}(\vec{p})$,  $a^c_{\sigma}(\vec{p})$ and their conjugates, whose commutation relations, obtained in \cite{Kazm6}, are
\be\label{coma}
\begin{aligned}
&\left[{a_{\sigma}(\vec{p}),a^{\dag}_{\sigma'}(\vec{p}\,')}\right]_+=
\left[{a^c_{\sigma}(\vec{p}),a^{c\dag}_{\sigma'}(\vec{p}\,')}\right]_+=
(2\pi)^32E_p\delta_{\sigma\sigma'}\delta(\vec{p}-\vec{p}\,'),\\
&\left[{a_{\sigma}(\vec{p}),a_{\sigma'}(\vec{p}\,')}\right]_+=
\left[{a^c_{\sigma}(\vec{p}),a^{c}_{\sigma'}(\vec{p}\,')}\right]_+=
\left[{a_{\sigma}(\vec{p}),a^{c\dag}_{\sigma'}(\vec{p}\,')}\right]_+=
\left[{a^c_{\sigma}(\vec{p}),a^{\dag}_{\sigma'}(\vec{p}\,')}\right]_+=0 ,
\end{aligned}
\ee
should be thought of as describing Dirac particles and their anti--particles in the asymptotic {\it in} and {\it out} regions. Of curse, the commutators of $a$'s with $b$'s all vanish. This follows from the fact that both $A_i$ and $\pi_{\perp}^i=\dot{A}_i$ have vanishing Dirac brackets with spinorial variables (see (\ref{AG}) and (\ref{Fpiperp})).
The Hilbert space carrying a representation of the commutation relations between the interaction picture fields, can now be simply constructed as the Fock space of $a$, $a^{\dag}$, $a^c$, $a^{c\dag}$, $b$, $b^{\dag}$. The asymptotic {\it in} and {\it out} states will inhabit this space. The non--interacting vacuum (i.e. the lowest energy state for the free Hamiltonian $\widehat{H}_0$), from which this Fock space is constructed, satisfies
\be
a_{\sigma}(\vec{p})|0\rangle=a^c_{\sigma}(\vec{p})|0\rangle=
b_{\lambda}(\vec{k})|0\rangle=0 .
\ee

 To proceed with this description, I shall use the well known formula for the perturbaive expansion of the $S$ matrix elements (the so called {\it Dyson series}):
\be\label{Dyson1}
S(1,2,\cdots \rightarrow  1',2',\cdots)=\sum_{N+0}^{\infty}\frac{(-i)^N}{N!}\int\cdots\int d^4x_1\dots d^4x_N
\langle 0|\dots a_{2'} \, a_{1'} \, T\left\{{:h(x_1):\dots :h(x_N):}\right\} \, a_1^{\dag} \, a_2^{\dag}\dots|0\rangle
\ee
(compare with (6.1.1) of \cite{Wein}). Here $1,2,\cdots$ denote the incoming particles, $1',2',\cdots$ are outgoing particles (these numbers are assumed to include the information about the momenta, spin projections, helicities and types of particles), $T\{\}$ is the time ordering, $h(x)$ is the interaction Hamiltonian density in the interaction picture and $: :$ denotes normal ordering operation (i.e. all the annihilation operators occur on the right of the creation operators). 
Using (\ref{Hcoul}) and (\ref{A0ost}), one obtains
\be\label{h1}
\ba
&h(x)=h_A(x)+h_C(x),\\
&h_A(x)=e\widehat{A}_{(I)j}(x)\l{\widehat{\ov{\psi}}_{(I)}\gamma^j\widehat{\psi}_{(I)}}\r(x),\qquad
h_C(x)=\frac{e^2}{2}\int d^4y\frac{\delta(x^0-y^0)}{4\pi|\vec{x}-\vec{y}|}
\l{\widehat{\ov{\psi}}_{(I)}\gamma^0\widehat{\psi}_{(I)}}\r(x)
\l{\widehat{\ov{\psi}}_{(I)}\gamma^0\widehat{\psi}_{(I)}}\r(y) .
\ea
\ee

\section{Examples}\label{Exam}

\subsection{Compton scattering}

The Compton scattering is the process in which a photon with initial four--momentum $k$ and polarization
\be
e_i(k)=\alpha_+e_{i,1}(k)+\alpha_-e_{i,-1}(k),\qquad |\alpha_-|^2+|\alpha_+|^2=1
\ee
and the electron with initial four--momentum $p$ and the projection of spin onto the third spatial axis $\sigma$ interact to produce the outgoing photon with four--momentum $k'$ and polarization
\be
e'_i(k')=\alpha'_+e_{i,1}(k')+\alpha'_-e_{i,-1}(k'), \qquad |\alpha'_-|^2+|\alpha'_+|^2=1
\ee
and the outgoing electron with four--momentum $p'$ and the spin projection $\sigma'$. The {\it circular polarization} of the photon corresponds to $\alpha_-=0$ or $\alpha_+=0$, whereas the {\it linear polarization} corresponds to $|\alpha_-|=|\alpha_+|$. 

Since the calculations are going to be lengthy, it is extremely useful to simplify the notation. I shall define
\be\label{definitions}
\ba
&f^{\dag}:=\alpha_+b^{\dag}_1(\vec{k})+\alpha_-b^{\dag}_{-1}(\vec{k}),\qquad
f':={\alpha'}_{+}^{*}b_1(\vec{k}\,')+{\alpha'}_-^*b_{-1}(\vec{k}\,'),\qquad
a^{\dag}:=a^{\dag}_{\sigma}(\vec{p}),\qquad 
a':=a_{\sigma'}(\vec{p}\,'), \\
&x_{l'}^{(+)}:=\int d\Gamma_{p''}e^{-ip''x}u_{l'\sigma''}(\vec{p}'')a_{\sigma''}(\vec{p}''),\qquad
x_l^{(-)}:=\int d\Gamma_{p''}e^{ip''x}v_{l\sigma''}(\vec{p}'')a^{c\dag}_{\sigma''}(\vec{p}''),\qquad
x_l:=x_l^{(+)}+x_l^{(-)}, \\
&x_{i}^{(+)}:=\int d\Gamma_{k''}e^{-ik''x}e_{i\lambda}(\vec{k}'')b_{\lambda}(\vec{k}''),\qquad
x_i^{(-)}:={x_i^{(+)}}^{\dag},\qquad
x_i:=x_i^{(+)}+x_i^{(-)}, \\
&M^j:=\gamma^0\gamma^j .
\ea
\ee
The interaction density in the interaction picture (\ref{h1}) can now be rewritten as
\be\label{h}
\ba
&h(x)=h_A(x)+h_C(x),\\
&h_A(x)=eM^j_{ll'}\, x_j\, x^{\dag}_l\, x_{l'}\, ,\qquad 
h_C(x)=\frac{e^2}{2}\int d^4y\frac{\delta(x^0-y^0)}{4\pi|\vec{x}-\vec{y}|}
x^{\dag}_l\, x_l\, y^{\dag}_{l'}\, y_{l'}\,  
\ea
\ee
and the Dyson series (\ref{Dyson1}) for the Compton scattering is
\be
S(f,a \rightarrow  f',a')=\sum_{N+0}^{\infty}\frac{(-i)^N}{N!}\int\cdots\int d^4x_1\dots d^4x_N
\langle 0|f' a' \, T\left\{{:h(x_1):\dots :h(x_N):}\right\} \, a^{\dag} f^{\dag}|0\rangle  .
\ee

The $N=0$ term gives
\be\label{forward}
\langle 0|f'a'a^{\dag}f^{\dag}|0\rangle=[a',a^{\dag}]_+[f',f^{\dag}]_-=
\l{2\pi}\r^62E_k2E_p\delta(\vec{k}-\vec{k}')\delta(\vec{p}-\vec{p}')
\delta_{\sigma\sigma'}
\l{{\alpha'}_+^*\alpha_+ +{\alpha'}_-^*\alpha_-}\r .
\ee
Assume now that we wish to calculate the probability amplitude of the scattering event in which the final momenta $\vec{k}'$, $\vec{p}\,'$ are at least slightly different than the initial momenta $\vec{k}$ and $\vec{p}$. This assumption is certainly allowable, since it is up to us probability of which we wish to calculate! Under this assumption, the delta functions vanish and one gets no contribution from the $N=0$ term. In fact, since the wave packets that describe the incoming beams of particles are usually not ideally localized in the momentum space, the {\it forward scattering} term (\ref{forward}) may contribute slightly to the measured values of the nontrivial scattering, but I will not discuss that kind of technical complications here.

For $N=1$ the two terms corresponding to $h_A$ and $h_C$ need to be evaluated
\be
-i\int d^4x\langle 0| f'a':h_A(x): a^{\dag}f^{\dag}|0\rangle
-i\int d^4x\langle 0| f'a'T\{:h_C(x):\} a^{\dag}f^{\dag}|0\rangle .
\ee
Note that $h_C$ needs to be time--ordered due to its nonlocal character (the time ordering of $h_A$ is not necessary). It is easy to verify that the first component vanishes. Hence, the Coulomb component needs to be calculated. However, this term also appears to vanish, which follows from the fact that
\be
\langle 0|a'\, :x^{\dag}_l\,x_l\,y^{\dag}_{l'}\,y_{l'}:\, a^{\dag}|0\rangle =0 .
\ee

\begin{figure}[t!]
\begin{minipage}{0.45\linewidth}
\includegraphics[trim = 35mm 150mm 0mm 35mm, clip, width=8cm]{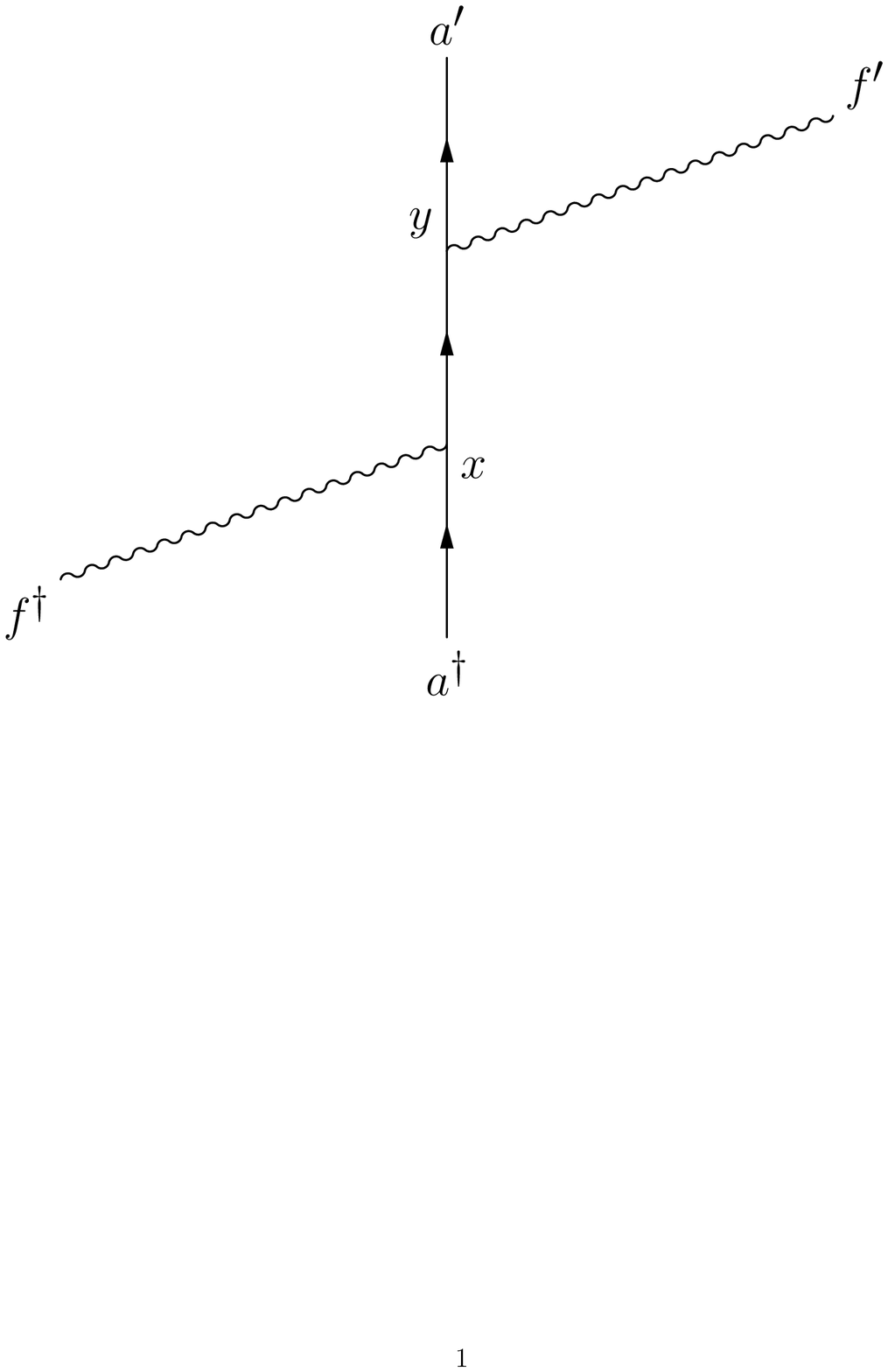}

\caption{The diagram describing the $\lambda$--term of (\ref{N=2fin}).}\label{lamb}
\end{minipage}
\begin{minipage}{0.45\linewidth}
\includegraphics[trim = 35mm 150mm 0mm 35mm, clip, width=8cm]{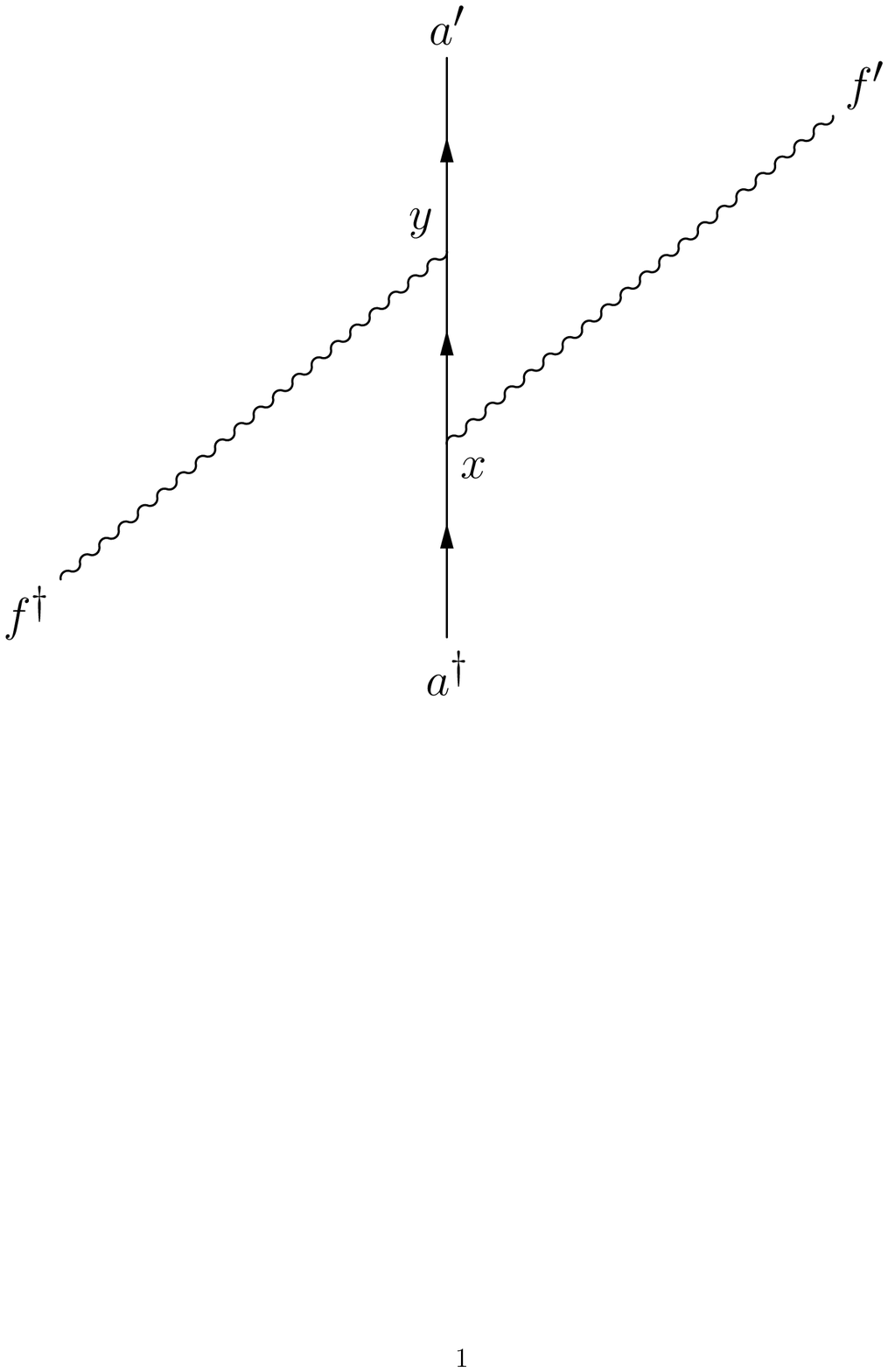}

\caption{The diagram describing the $\tilde{\lambda}$--term of (\ref{N=2fin}).}\label{tlamb}
\end{minipage}
\end{figure}

For $N=2$ the contribution is
\be\label{N=2}
\frac{(-i)^2}{2!}\iint d^4xd^4y
\langle 0| f'a'  \, T\left\{{:h(x): \, :h(y):}\right\} \, a^{\dag}f^{\dag}|0\rangle .
\ee
But
\be
\ba
&:h(x): \, :h(y):=\l{:h_A(x):+:h_C(x):}\r \l{:h_A(y):+:h_C(y):}\r \\
&= \ :h_A(x): \, :h_A(y):+:h_A(x): \, :h_C(y):+:h_C(x): \, :h_A(y):+:h_C(x): \, :h_C(y): \ .
\ea
\ee
From (\ref{h}) it is clear that the first term is proportional to $e^2$ and the remaining terms are of higher order in $e$. If we wish to find the correction to the $S$ matrix that is of lowest nontrivial order in the small coupling constant $e$, we should neglect all the terms but the first. A straightforward calculation then gives
\be\label{electfactor}
\ba
&\langle 0| f'a'  \, :h_A(x): \, :h_A(y): \, a^{\dag}f^{\dag}|0\rangle=
e^2M^i_{ll'}M^j_{nn'}
\langle 0| f'a' \, :x_ix^{\dag}_lx_{l'}:\, :y_jy^{\dag}_ny_{n'}: \, a^{\dag}f^{\dag}|0\rangle \\
&=e^2M^i_{ll'}M^j_{nn'}
\langle 0|f' x_i y_j f^{\dag} a'  \, :x^{\dag}_lx_{l'}:\, :y^{\dag}_ny_{n'}: \, a^{\dag}|0\rangle \\
&=e^2M^i_{ll'}M^j_{nn'}
\l{[f',x_i][y_j,f^{\dag}]+[f',y_j][x_i,f^{\dag}]+[f',f^{\dag}][x_i^{(+)}y_j^{(-)}]}\r
\langle 0|a' \, :x^{\dag}_lx_{l'}:\, :y^{\dag}_ny_{n'}: \, a^{\dag}|0\rangle
\ea
\ee
Here and below the bracket $[,]$ means the commutator for even variables and anti--commutator for odd ones (I drop the subscript $\mp$). Up to now, I have pulled out all the electromagnetic operators in terms of commutators. The last factor contains the free vacuum expectation value of the fermionic operators and it will now be computed
\be\label{fermfactor}
\ba
&\langle 0|a' \, :x^{\dag}_lx_{l'}:\, :y^{\dag}_ny_{n'}: \, a^{\dag}|0\rangle \\
&=[a',x^{\dag}_l][y_{n'},a^{\dag}]\langle 0| x_{l'}y^{\dag}_n|0\rangle-
[a',y^{\dag}_n][x_{l'},a^{\dag}]\langle 0|x^{\dag}_ly_{n'}|0\rangle+
[a',a^{\dag}]\langle 0|:x^{\dag}_lx_{l'}:\, :y^{\dag}_ny_{n'}: |0\rangle \\
&=[a',x^{\dag}_l][y_{n'},a^{\dag}] [ x^{(+)}_{l'},y^{(+)\dag}_n]-
[a',y^{\dag}_n][x_{l'},a^{\dag}] [x^{(-)\dag}_l,y^{(-)}_{n'}]+
[a',a^{\dag}]  [x^{(+)}_{l'},y^{(+)\dag}_n]     [x^{(-)\dag}_l,y^{(-)}_{n'}]  .
\ea
\ee
From (\ref{comb}) and (\ref{coma}) it follows that some of the brackets are proportional do the Dirac deltas between initial and final momenta
\be
[f',f^{\dag}]=
\l{2\pi}\r^32E_k\delta(\vec{k}-\vec{k}')
\l{{\alpha'}_+^*\alpha_+ +{\alpha'}_-^*\alpha_-}\r,\qquad
[a',a^{\dag}]=
\l{2\pi}\r^32E_p\delta(\vec{p}-\vec{p}')
\delta_{\sigma\sigma'} .
\ee
Recall that we are trying to calculate the probability amplitude of the scattering event in which the final momenta are different than the initial momenta (the discussion below the formula (\ref{forward})). Hence, all the terms that are proportional to $[f',f^{\dag}]$ or $[a',a^{\dag}]$ simply vanish. Then inserting (\ref{fermfactor}) into (\ref{electfactor}) and multiplying by $-1/2$ that was present in front of the integral in (\ref{N=2})  yields
\be\label{niechr}
\ba
&\langle 0| f'a'  \, :h_A(x): \, :h_A(y): \, a^{\dag}f^{\dag}|0\rangle=-\frac{e^2}{2}M^i_{ll'}M^j_{nn'} \\
&\l{
[f',x_i][y_j,f^{\dag}] [a',x^{\dag}_l][y_{n'},a^{\dag}] [ x^{(+)}_{l'},y^{(+)\dag}_n]-
[f',x_i][y_j,f^{\dag}] [a',y^{\dag}_n][x_{l'},a^{\dag}] [x^{(-)\dag}_l,y^{(-)}_{n'}]
}\right.\\
&+
\left.{
[f',y_j][x_i,f^{\dag}] [a',x^{\dag}_l][y_{n'},a^{\dag}] [ x^{(+)}_{l'},y^{(+)\dag}_n]-
[f',y_j][x_i,f^{\dag}][a',y^{\dag}_n][x_{l'},a^{\dag}] [x^{(-)\dag}_l,y^{(-)}_{n'}]
}\r .
\ea
\ee
In order to obtain the contribution to the $S$ matrix, this expression needs to be time--ordered and integrated over $d^4xd^4y$. Let as perform this task for the first two terms of (\ref{niechr}). They can be rewritten as
\be
\ba
-\frac{e^2}{2}\l{
\lambda_{ln}(x,y) [ x^{(+)}_l,y^{(+)\dag}_n]-
\lambda_{nl}(y,x) [ x^{(-)\dag}_l,y^{(-)}_n]
}\r,\qquad
\lambda_{ln}(x,y):=M^i_{l'l}M^j_{nn'}[f',x_i][y_j,f^{\dag}] [a',x^{\dag}_{l'}][y_{n'},a^{\dag}] .
\ea
\ee
Time ordering and integration then yields
\be\label{lambda}
\ba
-\frac{e^2}{2}\iint d^4xd^4y &\left\{{
\theta (x^0-y^0)\l{\lambda_{ln}(x,y) [ x^{(+)}_l,y^{(+)\dag}_n]-
\lambda_{nl}(y,x) [ x^{(-)\dag}_l,y^{(-)}_n]}\r
}\right.\\
+
&\left.{
\theta (y^0-x^0)\l{\lambda_{ln}(y,x) [ y^{(+)}_l,x^{(+)\dag}_n]-
\lambda_{nl}(x,y) [ y^{(-)\dag}_l,x^{(-)}_n]}\r
}\right\}\\
=
-e^2\iint d^4xd^4y &\left\{{
\theta (x^0-y^0)\lambda_{ln}(x,y) [ x^{(+)}_l,y^{(+)\dag}_n]
-\theta (y^0-x^0)\lambda_{nl}(x,y) [ y^{(-)\dag}_l,x^{(-)}_n]
}\right\}\\
=ie^2\iint d^4xd^4y & \lambda_{ln}(x,y) \triangle_{ln}(x,y),
\ea
\ee
where $\theta(t)$ is the step function taking value $0$ for $t<0$, $1/2$ for $t=0$ and $1$ for $t>0$, which can be conveniently represented as an integral
\be\label{theta}
\theta(t)=\lim_{\varepsilon\to 0^+} -\frac{1}{2\pi i}\int_{-\infty}^{\infty}\frac{e^{-its}}{s+i\varepsilon}ds ,
\ee
and
\be
\triangle_{ln}(x,y)=i\l{
\theta (x^0-y^0) [ x^{(+)}_l,y^{(+)\dag}_n]-\theta (y^0-x^0) [ y^{(-)\dag}_n,x^{(-)}_l]
}\r
\ee
is called the {\it propagator} for the Dirac field. 
Treating the last two terms of (\ref{niechr}) in the same way, a similar expression is obtained
\be\label{tlambda}
ie^2\iint d^4xd^4y  \tilde{\lambda}_{ln}(x,y) \triangle_{ln}(x,y),\qquad
\tilde{\lambda}_{ln}(x,y):=M^i_{l'l}M^j_{nn'}[f',y_j][x_i,f^{\dag}] [a',x^{\dag}_{l'}][y_{n'},a^{\dag}]
\ee
(note that $\tilde{\lambda}$ differs from $\lambda$ just by the interchange of $x_i$ and $y_j$).
Hence, the total contribution to the $S$ matrix that is proportional to $e^2$ is
\be\label{N=2fin}
ie^2\iint d^4xd^4y  \l{\lambda_{ln}(x,y)+\tilde{\lambda}_{ln}(x,y)}\r\triangle_{ln}(x,y),
\ee
with $\lambda$ and $\tilde{\lambda}$ given in (\ref{lambda}) and (\ref{tlambda}). The two terms of (\ref{N=2fin}) correspond to the Feynman diagrams FIG.\ref{lamb} and FIG.\ref{tlamb}.
It is however important to see that the usage of Feynman diagrams and postulating Feynman rules is not really necessary, but only useful. In principle one could simply work out the commutation relations in the Dyson series, just as I did. This procedure certainly does not require the presence of Poincar\'e invariance. 

Using (\ref{definitions}), the Dirac propagator can be calculated and recast in a convenient form
\be\label{dirprop}
\triangle (x,y)=\lim_{\varepsilon \to 0^+}\int \frac{d^4q}{(2\pi)^4}\frac{N(q)e^{-iq\cdot (x-y)}}{m^2-q^2-i\varepsilon},\qquad N(q)=u(q)u^{\dag}(q)=(m+\not\! q)\gamma^0,\quad \not\! q:=q_a\gamma^a .
\ee
Hrere $u(q)$ was assumed to be constructed according to the conventions adopted in \cite{Kazm5}, which are consequently used in this article. 

In order to calculate $\lambda_{ln}$ and $\tilde{\lambda}_{ln}$ explicitly, it is necessary to use
\be
\ba
&[a',x^{\dag}_{l'}]=e^{ip'x}u^*_{l'\sigma'}(p'),\qquad 
[y_{n'}a^{\dag}]=e^{-py}u_{n'\sigma}(p),\qquad
[f',x_i]=e^{ik'x}{e'}^*_i(k'),\qquad
[y_j,f^{\dag}]=e^{-iky}e_j(k), \\
&[f',y_j]=e^{ik'y}{e'}^*_j(k'),\qquad [x_i,f^{\dag}]=e^{-ikx}e_i(k).
\ea
\ee
This results follow straightforwardly from (\ref{definitions}). Inserting these results into the formulas for $\lambda$ and $\tilde{\lambda}$ given in (\ref{lambda}) and (\ref{tlambda}), one gets
\be\label{ltl}
\ba
&\lambda_{ln}(x,y)=e^{i(k'x-ky+p'x-py)}
\left[{\ov{u}(p'){\not\!e\,'}^*(k')}\right]_{\sigma'l}
\left[{\gamma^0{\not\!e}(k)u(p)}\right]_{n\sigma}, \\
&\tilde{\lambda}_{ln}(x,y)=e^{i(k'y-kx+p'x-py)}
\left[{\ov{u}(p'){\not\!e}(k)}\right]_{\sigma'l}
\left[{\gamma^0{\not\!e\,'}^*(k')u(p)}\right]_{n\sigma} ,\\
&\not\!e^*:=\gamma^i e^*_i .
\ea
\ee
inserting (\ref{ltl}) and (\ref{dirprop}) into (\ref{N=2fin}) one obtains the $e^2$ contribution to the $S$ matrix in the form
\be\label{p1}
\lim_{\varepsilon \to 0^+}
ie^2(2\pi)^4\delta^4(p+k-p'-k')\ov{u}(p')
\left\{{
\frac{{\not\!e'}^*(k')N(p+k)\gamma^0\not\!e(k)}{m^2-(p+k)^2-i\varepsilon}+
\frac{\not\!e(k)N(p-k')\gamma^0{\not\!e'}^*(k')}{m^2-(p-k')^2-i\varepsilon}
}\right\}
u(p) .
\ee
To arrive at this result, one should first perform the integrals w.r.t. $d^4x$ and $d^4y$, which will produce delta functions from the exponents of (\ref{ltl}). These delta functions can be used in the subsequent integration over $d^4q$. One ends up with one delta function that simply expresses the four--momentum conservation.

Note that the expressions in the denominators of (\ref{p1}), namely $m^2-(p+k)^2=-2p\cdot k$ and $m^2-(p-k')^2=2p\cdot k'$, never vanish, because the plane that is perpendicular to a null vector does not contain any time--like vectors. Therefore, the infinitesimal parameter $\varepsilon$ can be simply set to zero. Another useful observation is that $N(p+k)\gamma^0=m+\not\!p+\not\!k$ and $N(p-k')\gamma^0=m+\not\!p-\not\!k\,'$. Using all of these, the $e^2$ contribution to the $S$ matrix element for Compton scattering can be finally expressed as $(2\pi)^4\delta^4(p+k-p'-k')i\mathcal{M}$, where
\be
i\mathcal{M}=-ie^2\ov{u}(p')
\left\{{
{\not\!e'}^*(k')\frac{\not\!p+\not\!k+m}{2p\cdot k}\not\!e(k)+
\not\!e(k)\frac{\not\!p-\not\!k\,'+m}{-2p\cdot k'}{\not\!e'}^*(k')
}\right\}
u(p) .
\ee
This formula agrees with the one that is obtained by standard methods from Feynman rules. It has exactly the same form as the first formula of Chapter 5.5 of \cite{Peskin}.

\subsection{$e^+,e^-\longrightarrow \mu^+,\mu^-$ scattering}

Another exemplary calculation will concern the probability amplitude for the production of a pair of muon and anti--muon from scattering of electron and positron. The modification of the Lagrangian that would allow for inclusion of many kinds of fermions is straightforward: any term of the form $\ov{\psi}L\psi$, where $L$ is a matrix--differential operator, ought to be replaced by  $\ov{\psi}^{(r)}L\psi^{(r)}$, with $(r)$ labeling different kinds of particles. The two terms contributing to the interaction Hamiltonian are now
\be
\ba
&h_A(x)=h^e_A(x)+h^{\mu}_A(x), \\
&h_C(x)=\frac{1}{2}\int\frac{d^4y\delta(x^0-y^0)}{4\pi|\vec{x}-\vec{y}|}\l{\rho^e(x)+\rho^{\mu}(x)}\r\l{\rho^e(y)+\rho^{\mu}(y)}\r=
h_C^e(x)+h^{\mu}_C(x)+
\int\frac{d^4y\delta(x^0-y^0)}{4\pi|\vec{x}-\vec{y}|}\rho^e(x)\rho^{\mu}(y), \\
&\rho^e(x)=e\ov{\psi}^e(x)\gamma^0\psi^e(x),\qquad \rho^{\mu}(x)=e\ov{\psi}^{\mu}(x)\gamma^0\psi^{\mu}(x) .
\ea
\ee
Let $a^{\dag}$, $a^{c\dag}$ denote the creation operators of electron and positron and $b$, $b^c$ the annihilation operators of muon and anti--muon. These operators will have to be supplemented by additional indexes for the momenta and spin projections of the corresponding particles, but I shall skip this labels in the beginning. 

The $N=1$ term in the Dyson series is 
\be\label{N1}
-i\int d^4x\langle 0|  bb^c T\left\{{:h_A(x):+:h_C(x):}\right\}  a^{c\dag}a^{\dag}|0\rangle .
\ee
The first component including $h_A$ is equal to
\be
\langle 0|  bb^c :h^e_A(x):   a^{c\dag}a^{\dag}|0\rangle+
\langle 0|  bb^c :h^{\mu}_A(x):   a^{c\dag}a^{\dag}|0\rangle=0 .
\ee
The vanishing of these two terms follows from the fact that in the first one can commute $b^c$ to the right, without producing any non--vanishing anti--commutators, whereas in the second it is possible to commute $a^{c\dag}$ to the left. The Coulomb part of (\ref{N1}) is, however, nontrivial and reads
\be\label{coulcontr}
-i\iint d^4xd^4y\frac{\delta(x^0-y^0)}{4\pi|\vec{x}-\vec{y}|}
\langle 0|  bb^c T\left\{{:\rho^e(x)\rho^{\mu}(y):}\right\}   a^{c\dag}a^{\dag}|0\rangle .
\ee
This terms constitutes the contribution of order $e^2$ following from the $N=1$ term of the Dyson series. I shall return to this contribution later.

The $N=2$ term of the Dyson series is
\be\label{p2}
-\frac{1}{2}\iint d^4xd^4y
\langle 0|  bb^c T\left\{{\l{:h_A(x):+:h_C(x):}\r\l{:h_A(y):+:h_C(y):}\r}\right\}  a^{c\dag}a^{\dag}|0\rangle ,
\ee
but the only term that is proportional to $e^2$ is
\be
-\frac{1}{2}\iint d^4xd^4y
\langle 0|  bb^c T\left\{{:h_A(x)::h_A(y):}\right\}  a^{c\dag}a^{\dag}|0\rangle .
\ee
The remaining terms in (\ref{p2}) are of higher orders in $e$. The calculations that are similar to those performed in the case of Compton scattering can now be performed, which allow one to recast these expression as
\be\label{prediagram}
\ba
&i\iint d^4xd^4y \triangle_{jk}(x,y)\lambda^{jk}(x,y), \\
&\lambda^{jk}(x,y):=e^2M^J_{ll'}M^k_{nn'}
[x^{e\dag}_l,a^{c\dag}][x^e_{l'},a^{\dag}][b,y^{\mu\dag}_n][b^c,y^{\mu}_{n'}],
\ea
\ee
where again the short notation was used
\be
\ba
&x^{e\dag}_l:=\psi^{(e)\dag}_l(x),\qquad x^{e}_{l'}:=\psi^{(e)}_{l'}(x),\qquad 
y^{\mu\dag}_n:=\psi^{(\mu)\dag}_n(y),\qquad y^{\mu}_{n'}:=\psi^{(\mu)}_{n'}(y),\\
&a^{\dag}:=a^{\dag}_{\sigma}(\vec{p}),\qquad a^{c\dag}:=a^{c\dag}_{\sigma'}(\vec{p}\,'),\qquad
b:=b_{\lambda}(\vec{k}),\qquad b^c:=b^c_{\lambda'}(k'),
\ea
\ee
where now $\vec{p}$, $\sigma$ and $\vec{p}\,'$, $\sigma'$ are the momenta and spin projections of the ingoing electron and positron, whereas $\vec{k}$, $\lambda$ and $\vec{k}\,'$, $\lambda'$ are the momenta and spin projections of the outgoing muon and anti--muon. The propagator for the electromagnetic field
\be
\triangle_{jk}(x,y):=i\l{
\theta (x^0-y^0) [ x^{(+)}_j,y^{(-)}_k]-\theta (y^0-x^0) [ y^{(+)}_k,x^{(-)}_j]
}\r,
\ee
(the definition of $x_i^{(+)}$ and  $x_i^{(-)}$ was already given in (\ref{definitions})) is equal to
\be
\triangle_{jk}(x,y)=-\lim_{\varepsilon \to 0^+}\int\frac{d^4q}{(2\pi)^4}\frac{N_{jk}(q)e^{-iq\cdot (x-y)}}{q^2+i\varepsilon},\qquad
N_{jk}(q)=\delta_{jk}-\frac{q_jq_k}{|\vec{q}|^2} .
\ee

\begin{figure}[t!]
\begin{minipage}{0.40\linewidth}
\includegraphics[trim = 35mm 165mm 0mm 36mm, clip, width=8cm]{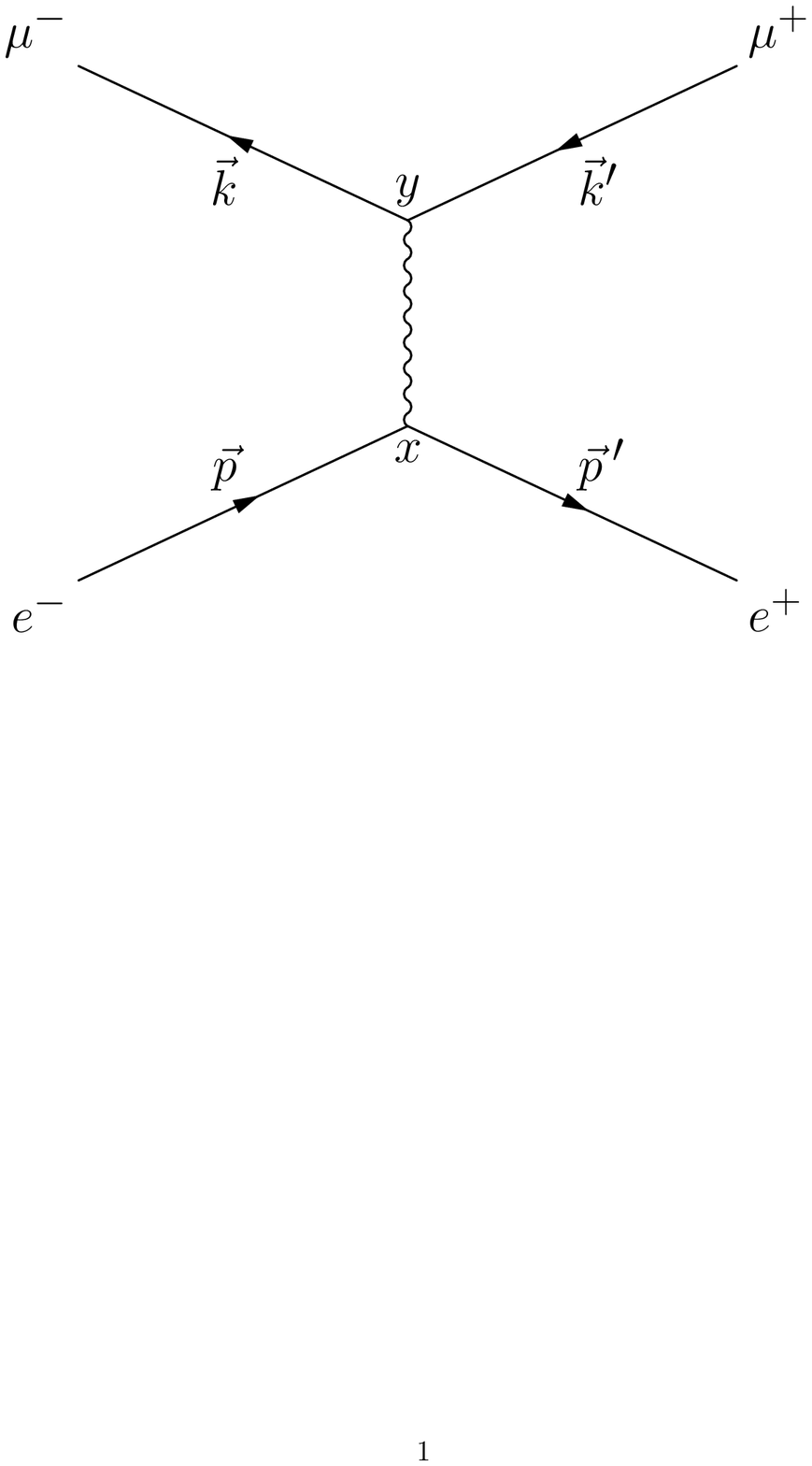}
%l b r t
\caption{The diagram corresponding to  (\ref{prediagram}).}\label{eemm}
\end{minipage}
\end{figure}

The expression (\ref{prediagram}) corresponds to the Feynman diagram FIG.\ref{eemm}.
If the calculations are continued, (\ref{prediagram}) acquires a form
\be\label{niecoul}
\ba
&(2\pi)^4\delta(k+k'-p-p')\frac{-ie^2}{q^2}\l{\delta_{jk}-\frac{\l{k_j+k_j'}\r\l{k_k+k_k'}\r}{|\vec{k}+\vec{k}'|^2}}\r\left[{\ov{v}(p')\gamma^ju(p)}\right]_{\sigma'\sigma}\left[{\ov{u}(k)\gamma^kv(k')}\right]_{\lambda\lambda'} ,\\
&q^2=(k+k')^2=(p+p')^2 ,
\ea
\ee
which differs from what would be obtained by the application of Lorentz--covariant Feynman rules to the diagram FIG.\ref{eemm}. It is now necessary to recall about the Coulomb term describing the $e^2$ contribution from $N=1$ term of the Dyson series. This term is given by (\ref{coulcontr}). Some simple manipulations employing the anti--commutation relations and invariance of the measure w.r.t. exchange of $x$ and $y$ allow to rewrite (\ref{coulcontr}) as
\be
-2ie^2\iint\frac{d^4xd^4y}{4\pi|\vec{x}-\vec{y}|}\delta(x^0-y^0)\theta(x^0-y^0)e^{-ix\cdot (p+p')}e^{iy\cdot (k+k')}
\left[{v^{\dag}(p')u(p)}\right]_{\sigma'\sigma}
\left[{u^{\dag}(k)v(k')}\right]_{\lambda\lambda'} .
\ee
Using the identities
\be
\frac{1}{4\pi|\vec{x}-\vec{y}|}=\int\frac{d^3q}{(2\pi)^3|\vec{q}|^2}e^{i\vec{q}(\vec{x}-\vec{y})},\qquad
\frac{\delta(x^0-y^0)}{4\pi|\vec{x}-\vec{y}|}=\int\frac{d^4q}{(2\pi)^4|\vec{q}|^2}e^{i q\cdot (x-y)},
\ee
and the fact that the step function obeys $\theta(0)=1/2$, one can finally rewrite the Coulomb contribution as
\be
(2\pi)^4\delta^4(k+k'-p-p')\frac{-ie^2}{|\vec{k}+\vec{k}'|^2}
\left[{v^{\dag}(p')u(p)}\right]_{\sigma'\sigma}
\left[{u^{\dag}(k)v(k')}\right]_{\lambda\lambda'}.
\ee
Adding this result to (\ref{niecoul}) and omitting $(2\pi)^4\delta^4(k+k'-p-p')$ we get ultimately
\be
i\mathcal{M}=\frac{ie^2}{q^2}\eta_{ab}
\left[{\ov{v}(p')\gamma^au(p)}\right]_{\sigma'\sigma}
\left[{\ov{u}(k)\gamma^bv(k')}\right]_{\lambda\lambda'},
\qquad q^2=(k+k')^2=(p+p')^2 .
\ee
This final result is clearly covariant and agrees with other references, e.g. \cite{Peskin}.

\section{Conclusions}
Given a Lagrangian formulation of a field theory, there exists an algorithmic procedure for finding all the constraints, the Hamiltonian and the commutation relations of all the fields with respect to the GDB. If there are first class constraints present, one can try a gauge invariant method of quantization such as BRST quantization (which was not discussed here) or, alternatively, one can eliminate gauge freedom by imposing gauge conditions. A requirement of consistency of gauge conditions with time evolution needs to be imposed and it may lead to additional constraints. If all the gauge freedom is eliminated, the remaining constraints are all second class and should be incorporated in the construction of GDB. Then, in order to pass to quantum theory, one should in principle seek for representation of the final commutation relations of important fields with respect to GDB in a Hilbert space.

All these steps were performed for the case of electrodynamics with fermions. The causal structure of space--time was not employed and, indeed, all these steps can still be performed if the electromagnetic interaction is replaced or supplemented by gravity. However, even in the case of electrodynamics, the resulting commutation relations appeared to be rather complicated and no obvious way of representing them in a Hilbert space was visible. In the simplest case of the free Dirac field, considered in the first article \cite{Kazm6}, it was possible to find explicit solution to the field equations. If the fields are constructed in such a way that they satisfy field equations automatically, then one does not need to bother about representing of their commutation relations with the Hamiltonian. What is more, the remaining commutation conditions between the functions that parametrize the exact solutions were sufficiently simple that the representation could be found for them. On the contrary, in the presence of electrodynamics the equations are to complicated to be solved explicitly. It is a way out of this problem in the case of electrodynamics, since the Hamiltonian decouples into the free and interaction part, the latter being proportional to a very small fine structure constant. This allows for the transition to the interaction picture, in which the equations and commutation relations are sufficiently simple. Then the perturbative quantization can be applied. In the case of gravity, no such obvious decoupling, which would lead to the simple interaction--picture equations, seems to be possible. On the other hand, the non--perturbative field equations are even more complicated then those of electrodynamics and hence the attempts to represent them, without any simplifications, in a Hilbert space seems hopeless. Certainly, one can decouple the metric tensor to the Minkowski part and the deviation, which is commonly done, and quantize only the deviation. Then the interaction picture can be obtained, but the background independence is sacrificed and the resulting theory is non--renormalizable. Hence, these problems are serious, and the lack of Poincar\'e invariance in the presence of gravity is certainly not an important obstacle. 

\section*{Acknowledgements}
This work was supported by grant N N202 287038.

\end{document}